\documentclass[useAMS,usenatbib]{mn2e}

\usepackage{graphicx}
\usepackage{amssymb}
\usepackage{lscape}

\def \oii {[O\,{\sc ii}]~}
\def \oiii {[O\,{\sc iii}]~}
\def \feii {[Fe\,{\sc ii}]~}
\def \feiii {[Fe\,{\sc iii}]~}

\title[The large area KX quasar catalogue]{The large area KX quasar catalogue:
  I. Analysis of the photometric redshift selection and the complete quasar
  catalogue\thanks{Based on observations made with ESO telescopes at the La
    Silla Paranal Observatory under programme IDs 083.A0360 and
    085.A0359}\thanks{Based on observations collected at the Centro
    Astron\'omico Hispano Alem\'an (CAHA) at Calar Alto, operated
    jointly by the Max-Planck Institut f\"ur Astronomie and the
    Instituto de Astrof\'isica de Andaluc\'ia (CSIC).}}

\author[Maddox et al.]
{Natasha Maddox\thanks{nmaddox@ast.uct.ac.za}$^{1,2}$, Paul C. Hewett$^3$,
Celine P\'{e}roux$^4$, Daniel B. Nestor$^5$, 
\newauthor Lutz Wisotzki$^2$
\vspace*{6pt}\\
$^1$Astrophysics, Cosmology and Gravity Centre (ACGC), Astronomy
Department, University of Cape Town, Private Bag X3,\\ 
7701 Rondebosch, Republic of South Africa\\
$^2$Leibniz-Institut f\"ur Astrophysik Potsdam (AIP), An der Sternwarte 16, D-14482
Potsdam, Germany \\
$^3$Institute of Astronomy, University of Cambridge, Madingley Road,
Cambridge CB3 0HA, UK \\
$^4$Observatoire Astronomique de Marseille Provence, Marseille, Laboratoire
d'Astrophysique de Marseille, UMR6110, \\
CNRS/Universit\'{e} de Provence, France \\
$^5$Department of Physics and Astronomy, University of California, Los
Angeles, CA 90095-1547, USA\\ }

\begin{document}


\maketitle

\begin{abstract}

The results of a large area, $\sim$\,600 deg$^2$, $K$-band
flux-limited spectroscopic survey for luminous quasars are presented. The survey
utilises the UKIRT Infrared Deep Sky Survey (UKIDSS) Large Area Survey
(LAS) in regions of sky within the Sloan Digital Sky Survey (SDSS)
footprint. The $K$-band excess (KX) of all quasars with respect to
Galactic stars is exploited in combination with a photometric
redshift/classification scheme to identify quasar candidates for
spectroscopic follow-up observations. The data contained within this
investigation will be able to provide new constraints on the fraction
of luminous quasars reddened by dust with $E(B-V)\le$ 0.5\,mag. The
spectroscopic sample is defined using the $K$-band, $14.0 \le K \le
16.6$, and SDSS $i$-band limits of $i=$ 19.5, 19.7 and 22.0 over sky
areas of 287, 150 and 196 deg$^2$, respectively. The survey includes
$>$\,3200 known quasars from the SDSS and more than 250 additional
confirmed quasars from the KX-selection. A well-defined sub-sample of
quasars in the redshift interval $1.0 \le z \le 3.5$ includes 1152
objects from the SDSS and 172 additional KX-selected quasars. The
quasar selection is $>95$~per~cent complete with respect to known SDSS
quasars and $>95$~per~cent efficient, largely independent of redshift
and $i$-band magnitude. The properties of the new KX-selected quasars
confirm the known redshift-dependent effectiveness of the SDSS quasar
selection and provide a sample of luminous quasars experiencing
intermediate levels of extinction by dust. The catalogue represents an
important step towards the assembly of a well-defined sample of
luminous quasars that may be used to investigate the properties of
quasars experiencing intermediate levels of dust extinction within
their host galaxies or due intervening absorption line systems.

\end{abstract}


\section{Introduction}\label{sec:introduction}

The Sloan Digital Sky Survey (SDSS, \citealt{York2000}), with some
ten thousand square degrees of simultaneous five-band imaging and
extensive spectroscopic follow-up observations has revolutionised
studies of optically-selected luminous quasars and active galactic
nuclei (AGN).

Optical selection of quasars is highly effective at identifying large
samples of quasars. Between the SDSS, the 2dF Quasar Redshift Survey
(2QZ, \citealt{Croom2004}) and the 2dF-SDSS LRG and QSO Survey (2SLAQ,
\citealt{Croom2009}), more than 150\,000 spectroscopically confirmed
quasars have been catalogued between redshifts $0<z<5.5$. The 2QZ and
2SLAQ exploit the ultraviolet (UV) `excess' shown by many quasars with
respect to Galactic stars, and blue quasars with $z<2.2$ are confirmed
with reasonable efficiency ($\sim$\,50~per~cent). However, at higher
redshifts, intergalactic absorption shortward of the Ly$\alpha$
emission line reduces the UV-excess signal. Quasars suffering from
even a small amount of dust extinction, whether intrinsic to the
quasar or due to intervening absorption systems, show reduced UV flux
at all redshifts.

The SDSS improves on the UV-excess selection by employing multi-colour
selection, choosing as quasar candidates objects with colours that are
not consistent with the stellar locus (\citealt{Richards2002}). This
has been shown to be effective at including quasars experiencing small
to moderate amounts of dust extinction (\citealt{Richards2003}) and
non-standard spectral energy distributions (SEDs,
\citealt{Hall2002}). However, the selection suffers from significant
incompleteness at redshifts $2.5<z<3.0$, coincident with the peak in
the space density of luminous quasars, where the SDSS colours of
quasars and stars are indistinguishable. More importantly, in the
context of the ability to undertake a census of objects experiencing
moderate amounts of dust extinction, the SDSS quasar survey is
flux-limited in the $i$-band. The use of the redder $i$-band
represents a significant improvement on previous surveys with flux
limits in the $B$- or $V$-bands. However, the significant reduction in
flux, even in the observed-frame $i$-band, due to the presence of dust
with $E(B-V)\ge$ 0.2\,mag means that the majority of any such
population of quasars falls below the survey flux limit, with the
fractional loss increasing rapidly with redshift.

Quasar selection at a variety of other wavelengths is also possible,
each with strengths and weaknesses. Radio wavelengths are unaffected
by dust obscuration, but only a small fraction ($\sim$\,10~per~cent)
of all quasars are radio-loud, and thus are not representative of the
general quasar population. Deep X-ray surveys are also less sensitive
to dust obscuration, but are incapable of detecting quasars surrounded
by large column densities of gas. As will be discussed in
Section~\ref{subsec:balfrac}, the fraction of broad absorption line
quasars (BALQSOs), which are faint in X-rays, may be as large as
40~per~cent at certain redshifts and luminosities.  The ROSAT Bright
Survey \citep{Schwope2000} covers a large fraction of the sky, but is
too shallow to allow an effective census of quasar activity. The
upcoming eROSITA \citep{Cappelluti2011} mission, an all-sky imaging
survey aiming to catalogue up to three million AGN, is promising but
spectroscopic follow-up observations will still be required to confirm
the quasar identifications.

Space-based long-wavelength infrared (IR) observations cover large areas to
interesting depths, but limited angular resolution and the
difficulty of distinguishing objects with significant AGN
activity from among the much more numerous and heterogeneous
starburst-dominated sources remains a challenge when attempting to
compile large catalogues of objects. Space-based near-infrared (NIR) surveys,
notably the Wide-field Infrared Survey Explorer (WISE;
\citealt{Wright2010}), offer enormous promise but currently
exploitation is at a very early stage.

Ground-basesd observations at NIR wavelengths hold several advantages over
those described above. First, the effects of dust obscuration are greatly
diminished in the NIR with respect to optical wavelengths. Second, large areas
of sky can be covered to magnitude depths rivalling optical
surveys. Finally, as will be shown in Section~\ref{subsec:KX}, all known types
of quasars exhibit a straightforward NIR excess, enabling their
selection irrespective of the details of the restframe UV through
optical SED.

The Two Micron All Sky Survey (2MASS, \citealt{Skrutskie2006}), with
observations in the $J$, $H$ and $K_s$ bands, essentially covers the
entire sky, but the bright magnitude limits restrict detections to
either the local Universe or the very brightest objects. The almost
complete UKIRT Infrared Deep Sky Survey (UKIDSS,
\citealt{Lawrence2007}) aims to cover 7500 deg$^2$ to $K=18.3$,
i.e. nearly three magnitudes deeper than 2MASS, and forms the basis
for the investigation of the quasar population described
here. Specifically, our goal is to undertake a survey of sufficient
sky area, reaching a flux limit adequate to provide new constraints on
the intrinsic frequency of quasars experiencing dust extinctions up to
$E(B-V)\simeq$ 0.5\,mag, as well as the presence of quasars with
intrinsically red UV through optical SEDs.  The sub-types of quasars
anticipated to be found in such a large-area, NIR-selected survey are
outlined in the following sub-sections.

\subsection{Red and reddened quasars}

Quasars exhibit red colours either due to an intrinsically red
continuum, or, because of the presence of dust along the line of
sight. In the case of dust extinction, the observed shape of the
quasar SED is altered due to the wavelength-dependent reduction of
flux, with shorter wavelengths more affected. The change of SED shape,
as well as the loss of flux, can both lead to quasars being excluded
from existing catalogues. The dust can either be intrinsic to the
quasar and host galaxy, or external to the system along the line of
sight; both cases lead to similar observational effects. The present
study is sensitive to both red and dust-reddened quasars provided the 
spectra still show evidence of broad
emission lines.

The debate regarding the fraction of dust-reddened quasars currently
missing from optical surveys is ongoing. Previous work by
\citet{Maddox2008}, as described in Section~\ref{subsec:KX}, estimated
the fraction of luminous dust-reddened quasars to be $<$ 20~per~cent,
although the area covered by that survey was relatively small. The
fraction is larger than the 15~per~cent estimated from the SDSS
quasars \citep{Richards2003}, but significantly less than the large
fractions claimed from other studies (e.g. \citealt{White2003}, for
example).

Other recent work searching for dust-reddened quasars has focused on
selection of candidates at wavelengths other than the strongly
affected optical. The FIRST--2MASS Red Quasar Survey
\citep{Glikman2007} selects radio-loud objects matched to 2MASS, and
then requires very faint optical magnitudes to be considered a red
quasar candidate for follow-up optical and NIR spectroscopy. The study
is hampered by the bright NIR magnitude limit of $K<14$ and highly
uncertain $E(B-V)$ estimates, making the estimates of the reddened
fraction of 25--60~per~cent also uncertain. Deeper NIR imaging is
required to properly sample the general quasar population instead of
only the very brightest subset.

\subsection{Broad absorption line quasars}

It is well established that BALQSOs show signs of intrinsic dust
reddening greater than that of the general quasar population
(\citealt{Reichard2003}, \citealt{Dai2008}, \citealt{Gibson2009}, for
example). Thus BALQSOs are underrepresented in the optical quasar
surveys within which they are identified. In addition to the dust
reddening, BALQSO SEDs are characterised by large sections of almost
completely absent flux, altering their photometric properties in a
non-uniform way.

Attempts have been made to determine the intrinsic fraction of BALQSOs
within the general quasar population, but these efforts are hampered
by the selection effects of the underlying catalogues on which the
studies are based, which affect the BALQSOs and nonBALs
differently. Recently, \citet{Allen2011} have performed extensive
analysis on the SDSS DR6 quasar catalogue, investigating the selection
algorithm and its effect on the determination of the intrinsic BALQSO
fraction. The Allen et al.  study found the intrinsic fraction of
BALQSOs, using the classical \textit{balnicity index}, (BI,
\citealt{Weymann1991}), to be as high as 41$\pm$5~per~cent. The study
also found evidence of evolution of the BALQSO fraction with redshift,
which would not be explained by simple orientation effects. However,
this result remains somewhat uncertain due to the large completeness
corrections dependent on redshift, colour and luminosity that affect
the parent SDSS quasar sample.

Constructing the quasar sample at NIR-wavelengths, less affected by
both the dust intrinsic to the BALQSOs and the loss of flux at UV and
optical wavelengths due to the absorption troughs, will greatly assist
the determination of the intrinsic BALQSO fraction.

\subsection{Quasars with dusty intervening absorbers}

There are many hundreds of high redshift ($z>2$) quasars that have
absorption signatures from large neutral gas column density
($N_{HI}>2\times 10^{20}$cm$^{-2}$) foreground objects imposed on
their spectra. The chemical composition of these Damped Lyman alpha
(DLA) intervening absorption systems has been extensively studied, but
their dust content remains poorly constrained.

The debate regarding the fraction of DLA invervening absorption
systems that remain undetected due to intrinsic dust causing the
background quasar to fall below the flux limit of optically selected
samples is ongoing.  Results from a radio-selected sample of quasars
indicate that optical selection underestimates the number of DLAs due
to dust bias by at most a factor of two \citep{Ellison2001}. However,
the radio sample is small and the constraints are not tight.

Employing a sample of probable DLAs selected via detection of
Ca\texttt{II} (H \& K) absorption in SDSS quasar spectra, dust
reddening in the underlying quasar spectra with $E(B-V)\sim$ 0.1
has been detected (\citealt{Wild2005}, \citealt{Wild2006}). Based on
the distribution of $E(B-V)$ values and the corresponding detection
probabilities, the fraction of systems missed due to the dust bias is
estimated to be as large as $\sim$\,40~per~cent, but still consistent 
with the estimates from the radio-selected study.

The Ca\texttt{II}-based absorption system investigation focussed on
relatively low redshifts, whereas, at higher redshifts $z>2$, it is
clear that the observed sample of DLAs shows no evidence for dust
reddening \citep{Frank2010}. Making progress towards establishing whether
a significant fraction (some tens of per cent) of DLA systems at high
redshifts also show evidence for modest amounts of dust extinction
requires a flux-limited quasar sample at NIR wavelengths, where the
bias against including such systems is greatly reduced.

\subsection{Quasars Selected at NIR wavelengths}

The current work is an extension of a previous programme described in
\citet{Maddox2008}. The goal is to exploit the $K$-band excess shown
by all quasars, irrespective of redshift and the amount of
dust-reddening, to construct a more complete quasar catalogue for
redshifts $1\le z\le 3.5$. Exploring areas of sky previously surveyed
by the SDSS greatly reduces the amount of new observations
required. The project is unique as it represents the first large-area
survey of quasar candidates selected by photometric redshift
techniques with uniform spectroscopic follow-up observations to
confirm the quasar identifications.

The outline of the paper is as follows. Section~\ref{sec:data}
describes the data used for the study, Section~\ref{sec:candselection}
details the selection of quasar candidates, Section~\ref{sec:obs}
describes the spectroscopic observations.  The KX Quasar Catalogue is
described in Section~\ref{sec:KXcatalogue}, while an evaluation of the
quality of the quasar selection algorithm is given in
Section~\ref{sec:photozeval}. Section~\ref{sec:kxproperties} describes
some of the properties of the KX-selected quasars, and a discussion of
the results is given in Section~\ref{sec:discussion}. For the
interested reader, the completeness of the candidate list with respect
to the input data and how it affects the effective area of the survey
is discussed in the Appendix. Concordance cosmology with $H_{0} = 70$
km s$^{-1}$ Mpc$^{-1}$, $\Omega_{m} = 0.3$ and $\Omega_{\Lambda} =
0.7$ is assumed throughout.  Magnitudes on the Vega system are used
throughout the paper, as the UKIDSS magnitudes are based on this
system. The SDSS AB-magnitudes are converted to the Vega system using
the relations: $u=u_{\rm AB}-0.93$, $g=g_{\rm AB}+0.10$, $r=r_{\rm
  AB}-0.15$, $i=i_{\rm AB}-0.37$, and $z=z_{\rm AB}-0.53$, as
described in \citet{Hewett2006}. Unless otherwise specified, the PSF
magnitudes are used for the SDSS $ugriz$ bands, and the aperture
corrected \texttt{aperMag3} magnitudes are used for the WFCAM $YJHK$
bands.

\section{Input Optical and NIR Data}\label{sec:data}

\begin{figure*}
\resizebox{\hsize}{!}{\includegraphics{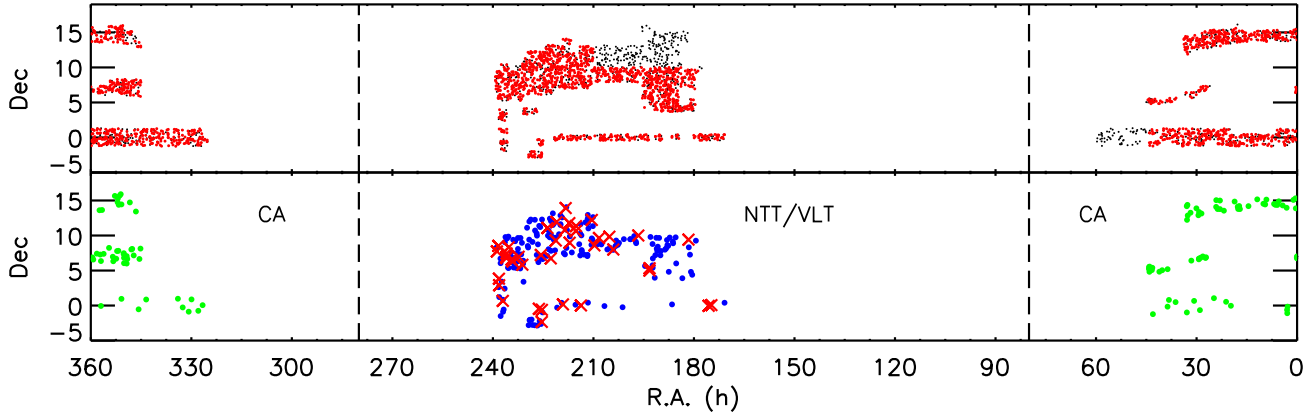}}
\caption{(Top) Area of sky where UKIDSS LAS DR4 was crossmatched to
  SDSS DR7 (small black dots). The red dots show the area of sky from
  which KX-survey candidates were chosen. (Bottom) Sky coverage of the
  observations. Green dots indicate objects observed by the Calar Alto
  telescopes, blue dots are objects observed with the NTT, and red
  crosses are VLT objects. The NTT/VLT region has fainter $i$- and
  $K$-band magnitude limits than the CA region.}
\label{fig:coverage}
\end{figure*}

The optical data, including photometry and spectroscopy, are taken
from the SDSS Data Release 7 (DR7, \citealt{Abazajian2009}). The NIR
photometric data are provided by the UKIDSS Data Release 4 (DR4) Large
Area Survey (LAS), which contains $\sim$\,900~deg$^2$ of coverage in
the $YJHK$ filter set described in \citet{Hewett2006}. The throughput
of the $ugrizYJHK$ filters can be seen in Fig.~7 of
\citet{Chiu2007}. The UKIDSS LAS reaches 5$\sigma$-depths of $Y=20.3$, $J=19.7$,
$H=19.0$, and $K=18.3$, i.e. approximately three magnitudes deeper
than 2MASS.  As the LAS lies within the SDSS footprint, each detected
object will potentially possess $ugrizYJHK$ photometry plus extensive
morphological information and, for a subset of targets, SDSS
spectroscopic observations.

The area of sky covered by the overlap of SDSS DR7 and UKIDSS LAS DR4,
along with the regions from which KX-survey candidates were chosen,
are shown in the top panel of Fig.~\ref{fig:coverage}. The data are
located in two distinct regions of the sky, hereinafter referred to as
the NTT/VLT region and the CA region, as labelled in the figure.

The reference list of spectroscopically confirmed SDSS quasars used
extensively in this work is the published SDSS DR7 quasar catalogue of
\citet{Schneider2010}, which contains 105\,783 unique objects,
supplemented by the list compiled by
\citet{Hewett2010}\footnote{http://www.sdss.org/dr7/products/value\_added/index.html},
containing 107\,194 entries. Object redshifts are taken from the
\citet{Hewett2010} catalogue and the sample of 107\,194 quasars/AGN is
herein referred to as the enhanced SDSS DR7 quasar catalogue.

\section{Quasar Candidate Selection}\label{sec:candselection}

This section details the principles and techniques used to construct
the list of quasar candidates for further spectroscopic
observations. Analysis of the performance of the candidate selection
is presented in Section~\ref{sec:photozeval}, and details regarding
the input catalogues that affect the completeness of the resulting
candidate list are presented in Appendix~\ref{app:catcompl}.

\subsection{$K$-band Excess of Quasars}\label{subsec:KX}

\begin{figure}
\resizebox{\hsize}{!}{\includegraphics{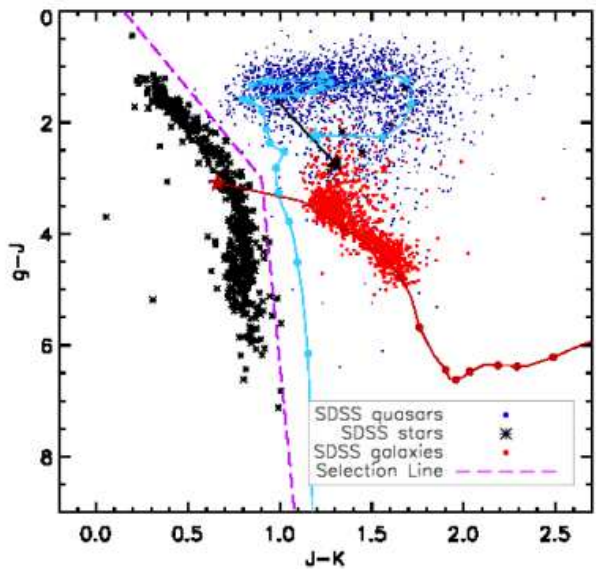}}
\caption{$g-J$ {\it vs} $J-K$ ($gJK$) plot showing the locations of the
  different subpopulations of objects. Spectroscopically confirmed SDSS stars
  (black asterisks), quasars (blue dots) and galaxies (red dots), restricted
  to $K\le 16.7$ are shown. The galaxies have been restricted to $z\ge0.15$
  for clarity. The stars and quasars are effectively separated at all
  redshifts, and remain on either side of the purple dashed selection line.
  The blue track identifies the locus of colours for a model unreddened quasar
  between $0<z<5$, and the red track is for an elliptical galaxy at $0<z<2$.
  The large stars on the tracks indicate $z=0$, and filled dots along the
  tracks are spaced at $\Delta z=0.2$ intervals.  The black arrow shows the
  movement of a quasar at $z=1$ if subjected to dust reddening of
  $E(B-V)=0.25$\,mag. The overlap of low redshift galaxies and high redshift or
  reddened quasars can be disentangled by utilising morphological information,
  as low redshift galaxies are, in general, resolved.}
\label{fig:gJK_sdss}
\end{figure}

The principle of utilising the $K$-band excess of quasars with respect
to Galactic stars for quasar selection was introduced in
\citet{Warren2000}. The key features of the $K$-band excess (KX)
quasar selection are illustrated in Fig.~\ref{fig:gJK_sdss}.  For this
figure, spectroscopically confirmed quasars, galaxies and stars were
extracted from the SDSS database and matched to the UKIDSS LAS. Each
population was restricted to $K\le 16.7$, and the galaxies were
restricted to $z>0.15$, for clarity within the plot. The quasars span
the redshift interval $0.2<z<5.0$.

The quasars show a clear $K$-band excess in their $J-K$ colours with
respect to stars at all redshifts and, importantly, the effect of dust
reddening increases the separation, as shown by the black arrow in the
figure.  The overlap within the colour-space between quasars and low
redshift galaxies can be resolved using morphological information when
targetting quasars with redshifts $z\ge1$ such that host galaxies are
not detected in the images.

There have been a number of studies employing KX selection of quasar
candidates over small ($<$\,1 deg$^2$) areas (for example,
\citealt{Sharp2002}, or more recently, \citealt{Smail2008}) but our
primary goal is to utilise the technique to examine the properties of
intrinsically luminous quasars where a large survey ares is required
due to their low surface density on the sky (tens of objects per
square degree).

The first large area study employing KX selection of quasars was
undertaken by \citet[][hereinafter `the Pilot KX Survey']{Maddox2008},
which covered $\simeq$\,13~square degrees. Aside from a $K$-band
magnitude restriction, the only criterion required for selection as a
KX candidate was a position on the $gJK$ plane rightward of a
selection line similar to that shown in Fig.~\ref{fig:gJK_sdss}.  In
addition to recovering known SDSS quasars, a number of quasars not
selected by the SDSS algorithm were observed. These objects included
low-redshift quasars with substantial contamination from their host
galaxies, high-redshift BALQSOs, and quasars at $2.5<z<3.0$, where the
SDSS selection algorithm is highly incomplete. The aim of the current
study is to build on the information gained from the Pilot KX Survey
to further improve the candidate selection and increase the area
covered by more than an order of magnitude.

\subsection{Photometric Redshifts}\label{subsec:photoz}

Photometric redshift (Photo-$z$) algorithms aim to determine the
identification and redshift of an object based primarily on broad band
photometry, without the use of time consuming spectroscopy. Much
effort has been focused on Photo-$z$ development in recent years due
to the large area, deep, multi-wavelength imaging surveys that have
become available or are in the final planning stages. The large number
of objects and their faint magnitudes make follow-up spectroscopy
unrealistic, leaving Photo-$z$ as the only viable alternative for
object classification and redshift determination.

In their simplest form, Photo-$z$ algorithms attempt to determine both
the identification and redshift of an input object by comparing the
observed photometry to synthetic photometry from a set of either model
or template SEDs.  The identification and redshift are found by
determining the best match between the observed and synthetic
photometry. Although Photo-$z$ is conceptually straightforward,
achieving accurate, precise and reliable results is difficult.

Several different Photo-$z$ algorithms are freely available, with
varying levels of complexity, such as \textit{Hyperz},
\citep{Bolzonella2000}, and EAZY, \citep{Brammer2008}. Most
concentrate on identifying galaxies, exploiting in particular the
distinctive 4000\AA-break spectral feature characteristic of old
stellar populations.

There are studies focusing on Photo-$z$ for quasars
(\citealt*{Hatziminaoglou2000}, \citealt{Richards2009} and references
therein) but the results are generally not as good as those for
galaxies due to the lack of strong spectral features in the power-law
quasar spectrum over large ranges of redshift. At $z<1$, the quasar
host galaxy can also complicate Photo-$z$ identification by
contributing significant galaxy-flux to the quasar SED.

\subsection{The NIR Photo-$z$ Algorithm}

The Photo-$z$ algorithm used to select candidates for the present
study (hereinafter referred to as the NIR Photo-$z$) was custom-made
for this project and is relatively simple when compared to the more
complex Photo-$z$ algorithms recently released. Within the NIR
Photo-$z$, the `identification' and `redshift' are determined using
straightforward $\chi^2$ minimisation between the observed photometry
and synthetic photometry derived from model and template SEDs and the
$ugrizYJHK$ passbands. However, the NIR Photo-$z$ holds two advantages
over previous quasar Photo-$z$ efforts. The first is the use of
optical--NIR photometry afforded by the overlapping areas of SDSS and
UKIDSS.  As described in Section~\ref{subsec:KX}, the combination of
optical and NIR colours allows the effective discrimination between
stars and quasars at all redshifts due to the excess of $K$-band flux
shown by quasars.

The second advantage is the high-quality quasar models employed.
Eleven model quasars, with a variety of SED properties were used
within the NIR Photo-$z$. The base-model is the unreddened, `standard'
model, quasar described in Section~2.4 of \citet{Maddox2008}. Seven
reddened models were created, taking the base-model, reddened with
increasing amounts of SMC-type dust covering the interval
\mbox{$0.1\le E(B-V)\le1.0$}.  In addition, models with emission
line strength twice that of the standard model, combined with both the
standard and a redder continuum slope ($\alpha=-0.54$ and
$\alpha=-1.04$, respectively, for $\lambda<2750$\,\AA), as well as a particularly
blue model ($\alpha=-0.04$) with no emission lines were incorporated
into the NIR Photo-$z$. Experiments were conducted to ascertain
whether a larger range of quasar SED properties improved the
performance of the redshift determinations but no significant gain was
found from the addition of further models.

To identify morphologically compact galaxies and remove them from
further consideration, seven galaxy templates, spanning E through Scd
types, were used. The E through Sc templates are from
\citet{Mannucci2001}, while the Sbc and Scd galaxy values are from
\citet{CWW1980}. The quasar models and galaxy templates were employed
to generate synthetic colours for redshift increments of $\Delta
z=0.025$, over the intervals $0.0<z\le5.0$ for quasars and
$0.0<z\le3.6$ for galaxies, although due to the relatively bright
magnitude limits, galaxy templates at redshifts of $z_{phot}>1.0$ were
never found to be chosen as the best solution for any objects.

Star templates, from the Bruzual-Persson-Gunn-Stryker atlas, were also
incorporated. The atlas is unpublished, but can be found at
http://www.stsci.edu/hst/observatory/cdbs/bpgs.html.

\subsection{Final Quasar Candidate Selection}\label{subsec:candsel}

The following steps are involved in creating the final quasar
candidate list from the initial optical and NIR photometry
data. First, objects with stellar morphological classification as
determined by the UKIDSS data reduction pipeline, and have detections
in all four $YJHK$ passbands covering the area in the sky shown in
Fig.~\ref{fig:coverage} are selected from the LAS DR4 database via the
WFCAM Science Archive (WSA, \citealt{Hambly2008}) SQL query
interface. The resulting data from this query are crossmatched to the
SDSS DR7 database (\texttt{PhotoObjAll}, \texttt{SpecObjAll} and
\texttt{Photoz} tables). Only the nearest neighbour within a pairing
radius of 1$\arcsec$ was retained, as the astrometric consistency
between the SDSS and UKIDSS is better than $0.5\arcsec$
\citep{Chiu2007}. The crossmatched UKIDSS LAS and SDSS data produced a
list of objects that possess $ugrizYJHK$ magnitudes, morphology
information, and for subsets of objects, SDSS spectra and SDSS
Photo-$z$ results. Objects detected in the UKIDSS NIR bands but not in
the SDSS optical bands, and thus have very red colours, are not
included in the current study.

The sample is initially restricted to $K\le17.0$, as extending to
fainter magnitudes not only increases the photometry errors (and thus
reduces the effectiveness of the NIR Photo-$z$ routine), but also
makes the optical spectroscopic follow-up more difficult. In addition,
with a brighter $K$-band limit, objects that are faint in the optical
must have very red optical--NIR colours. This puts strong constraints
on the range of possible identifications available, even if no
identification is made from the spectrum.

Candidates lying leftward of the selection boundary in the
$gJK$-space, illustrated in Fig.~\ref{fig:kx_gJK}, are discarded. Less
than one per cent of the SDSS DR7 quasars crossmatched to UKIDSS lie
to the left of the selection line. The small percentage of quasars
removed by application of the criterion will, by definition, be blue
in $J-K$, and are thus extremely unlikely to possess red restframe
optical SEDs. The benefit of the KX-selection is evident from the
elimination of the vast majority of Galactic stars with more than 90
per cent of all cross-matched UKIDSS-SDSS stellar objects removed in a
single step.

Extensive effort has been devoted to the assignment of reliable
photometric redshifts to low-redshift galaxies within the SDSS
(\citealt{Csabai2003}; \citealt{Oyaizu2008};
\citealt{Cunha2009}). Objects identified as galaxies with an SDSS
Photo-$z$ in SDSS DR7 were removed next.

At this stage, all cross-matched objects with properties consistent
with main-sequence stars or resolved galaxies have been removed,
leaving just apparently stellar objects to the right of the selection
boundary in $gJK$-space. The list of such objects with $K\le17.0$ is
run through the NIR Photo-$z$ routine, producing lists of probable
galaxies, stars, and quasar candidates. The probable galaxies and
stars are removed from further consideration. The NIR Photo-$z$ code
was tested on the full Pilot KX Survey catalogue of objects to ensure
the robustness of the quasar, star and galaxy classifications. The
performance of the routine for identifying all three classes of
objects was found to be very good.

The list of quasar-candidates constitutes the primary sample to be
used to investigate the properties of objects identified via the use
of NIR-photometry that eluded the SDSS quasar selection. To ensure the
most efficient use of follow-up spectroscopic observations, two
subsets of candidates are flagged as `not to be followed-up
spectroscopically'. First, objects with SDSS spectra are removed, as
their identification is already known.  Second, objects with SDSS
spectroscopic quasar-target flags set, but no spectroscopic
observations, are also removed. This restriction is discussed further
in Section~\ref{subsec:TARGETs}. Both sub-samples form part of the
final quasar catalogue.

The large number of relatively low-redshift quasar candidates are
removed by restricting the sample to $z_{phot}\ge 1.0$.  The decision
ensures that potential complications arising from quasar host galaxies
do not impact on the completeness of the flux-limited NIR-selected
quasar sample. At low redshifts, not only can the colours of the
candidates be altered by the combination of quasar+host galaxy flux,
but the $K$-band flux from the galaxy can artificially boost the
quasar+host system brighter than the flux limit.  These issues can be
problematic at significant ($z>0.3$) redshifts for a $K\le17.0$ sample
(\citealt{Maddox2006}). Most importantly, given the restriction of the
cross-matched UKIDSS-SDSS sample to `stellar' objects, the
$z_{phot}\ge 1.0$ selection ensures that genuine quasars are not
systematically excluded due to the visibility of their host galaxies
in the NIR images. From a practical perspective the application of the
$z_{phot}\ge 1.0$ criterion ensures that the resulting sample of
quasars is dominated by relatively high redshift objects which possess
longer path-lengths for the occurence of intervening absorption
systems.

\begin{figure}
\resizebox{\hsize}{!}{\includegraphics{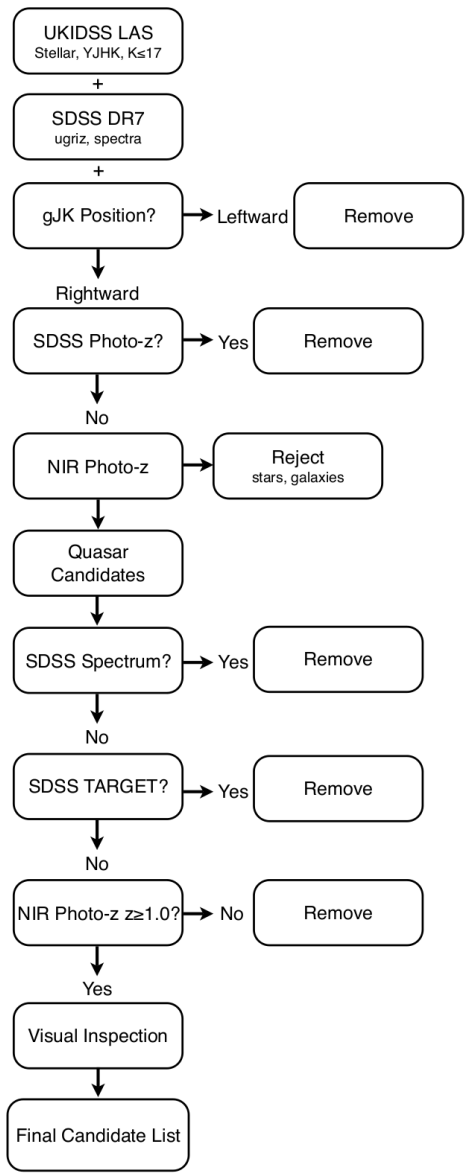}}
\caption{Flow-chart of the steps involved in the quasar candidate selection.}
\label{fig:chart}
\end{figure}

Finally, after all of these restrictions have been implemented, the
SDSS images of the quasar candidates are visually inspected to remove
objects whose photometry has been compromised, for example, by nearby
bright stars. There are a number of these in the survey area, and
their removal reduces the effective area by a small percentage, as
discussed in the Appendix. A flow-chart describing the quasar
candidate selection procedure is shown in Fig.~\ref{fig:chart}.

\section{New Spectroscopic Observations}\label{sec:obs}

In order to maximise the dynamic range of the survey in absolute
magnitude at fixed redshift, while ensuring adequate sampling of the
brightest quasars, a `wedding cake' survey geometry is adopted. Thus,
the entire survey area is complete to the brightest limits, with a
smaller sub-region complete to fainter magnitudes and the smallest
sub-region complete to the faintest limits.

\begin{table*}
\caption{Observing log for the quasar candidates}
\label{tab:allobs}
\centering
\scriptsize{
\begin{tabular}{ccccccccccccc}\\ \hline

Date & Telescope & Spectrograph & Grism & Slit Width (") & Wavelength Range
(\AA) & Resolution (R) \\
\hline
11--15 May 2009 & NTT & EFOSC2 & Grism \#11 & 1.0, 1.2 & $3320<\lambda<7500$ &
816--1845 \\
May--June 2009 & VLT & FORS & GRIS\_300V+10 & 1.6 & $4070<\lambda<9760$ &
1220--2930 \\
19--27 Aug 2009 & CA 2.2m & CAFOS & blue-200 & 1.5, 2.0 & $3750<\lambda<8870$ &
710--1830 \\
19--27 Aug 2009 & CA 2.2m & CAFOS & green-200 & 1.5, 2.0 &
$4790<\lambda<10000$ & 1040--2180 \\
31 Aug -- 02 Sept 2009 & CA 3.5m & MOSCA & green-500 & 1.5 &
$4080<\lambda<9700$ & 1450--3450 \\
April--July 2010 & VLT & FORS & GRIS\_300V+10 & 1.3 & $4080<\lambda<9700$ &
1120--2670 \\ \hline
\end{tabular}
}
\end{table*}

Observations of the survey area utilised four telescopes over three
observing semesters between 2009 May -- 2010 July. As the candidates
span six magnitudes in the $i$-band, smaller telescopes were used for
the optically bright objects and an 8-metre class telescope was
employed for the much less numerous NIR-bright but optically faint
objects.

Observations of the optically bright candidates located at $170<$ RA
$<240$ (NTT/VLT region) were carried out with the EFOSC2 spectrograph
\citep{Buzzoni1984} on the New Technology Telescope (NTT) at La Silla
observatory, and those at RA $>325$ or RA $<60$ (CA region) were
observed with the CAFOS spectrograph on the 2.2m telescope and MOSCA
on the 3.5m telescope at Calar Alto observatory.

The optically faint candidates in the NTT/VLT region were observed
with the FORS instrument on the European Southern Observatory (ESO)
Very Large Telescope (VLT) UT1 in service mode. The fainter candidates
were only observed in this region. Fig.~\ref{fig:coverage} shows the
location of the observed candidates and Table~\ref{tab:allobs} lists
the details of the observations. Moderate resolution spectra
($700<R<3000$) were obtained, generally sufficient for object
identification. In total, 324 candidates were observed, and the total
area of all the regions surveyed is 633 square degrees. The effective
area surveyed, accounting for area missed due to bad photometry, among
other effects, is 567.0 square degrees, and is described in the
Appendix.

\subsection{Magnitude and Redshift Limits}\label{subsec:completelimits}

The completeness limits for the entire area are $K\le16.5$ and $i\le19.5$.
For the NTT/VLT region at RA $<$ 210, the completeness limits are $K\le16.5$
and $i\le19.7$, and for RA $>$ 210 (hereinafter referred to as the NTT/VLT
Deep region), they are $K\le16.6$ and $i\le22.0$. The increased $i$-band depth
at RA $>$ 210 is because more faint targets were observed by the VLT within
this area. It is worth noting that this faint $i$-band magnitude limit is
nearing the detection limit of the SDSS photometry, and the NTT/VLT
Deep region is still nearly 200 square degrees in size. 

\begin{figure}
\resizebox{\hsize}{!}{\includegraphics{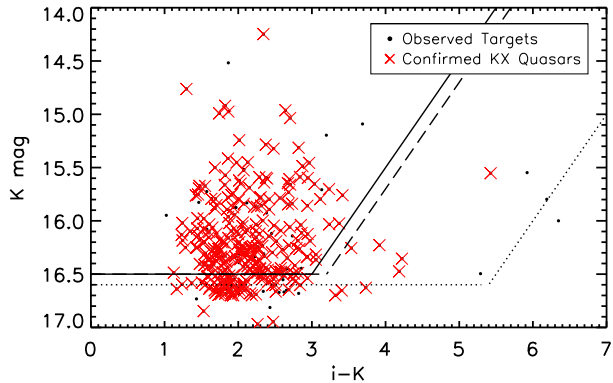}}
\caption{Colour-magnitude diagram for the observed candidates (black
  dots) and the spectroscopically confirmed KX-selected quasars (red
  crosses). The dotted lines show the region of colour-magnitude space
  to which the deepest area is complete, the dashed line shows the
  area complete to $K\le16.5$ and $i\le19.7$. Objects brighter than
  $K\le16.5$ with $i-K$ bluer than the solid diagonal line are
  considered as part of the complete catalogue which covers the entire
  area surveyed.}
\label{fig:ivsK}
\end{figure}

The $K$-band magnitude and $i-K$ colour limits are shown in
Fig.~\ref{fig:ivsK}, and a summary of the three regions with their
respective sizes in square degrees and magnitude limits is listed in
Table~\ref{tab:areas}. Some, but not all, of the available targets
with magnitudes fainter than the limits listed in
Table~\ref{tab:areas} were observed. These objects are included in the
full KX-selected quasar catalogue described in
Section~\ref{sec:KXcatalogue}.

The bright magnitude limits are $i\ge16.0$ and $K\ge14.0$. Only one
candidate is removed by these restrictions, and it is very unlikely
that a candidate with $K\le14.0$ is a quasar at $z\ge1.0$, as quasars
of this bright luminosity are intrinsically very rare. Inspection of
the SDSS DR7 quasar catalogue crossmatched to UKIDSS DR7 shows that of
20148 quasars with $1.0\le z\le3.5$, there are only 11 with
$K\le14.0$.

The redshift range over which the catalogue is complete is $1.0\le
z\le3.5$. The low redshift cutoff was implemented in the candidate
selection, and the high redshift limit is determined from the
completeness investigation in Section~\ref{subsec:completeness}, where
it is found that the completeness of the NIR Photo-$z$ selection drops
rapidly at $z>3.5$.

\begin{table}
\centering
  \caption{\label{tab:areas}Size and magnitude limits of the regions
    covered. The areas listed are total areas, not effective areas.}
\begin{tabular}{lccc} \hline
Region & Area & $i$-band & $K$-band \\
 & (deg$^2$) & Range & Range \\ \hline
CA & 286.6 & $16.0\le i\le19.5$ & $14.0\le K\le16.5$ \\
NTT/VLT & 150.1 & $16.0\le i\le19.7$ & $14.0\le K\le16.5$ \\
NTT/VLT Deep & 196.4 & $16.0\le i\le22.0$ & $14.0\le K\le16.6$ \\ \hline
\end{tabular}
\end{table}

\subsection{Data Reduction and Object Classification}\label{subsec:obsresults}

Reduction of the spectra was performed using standard
IRAF\footnote{IRAF is distributed by the National Optical Astronomical
  Observatories, which are operated by the Association of Universities
  for Research in Astronomy, Inc., under cooperative agreement with
  the National Science Foundation} tasks. The spectra were flux
calibrated from observations of standard stars taken at the beginning
or end of each observing night with appropriate grism and slit width
combinations. The standard star observations were only sufficient to
determine a relative flux calibration (i.e. the shape of the spectra
were corrected but the normalisation was only approximate). Comparison
between the computed spectrophotometric magnitudes and the known
optical magnitudes confirmed that an offset correction was required,
but that the overall shape of the flux calibrated spectra was
consistent with the optical colours to within a few tenths of a
magnitude.

\begin{figure}
\resizebox{0.9\hsize}{!}{\includegraphics{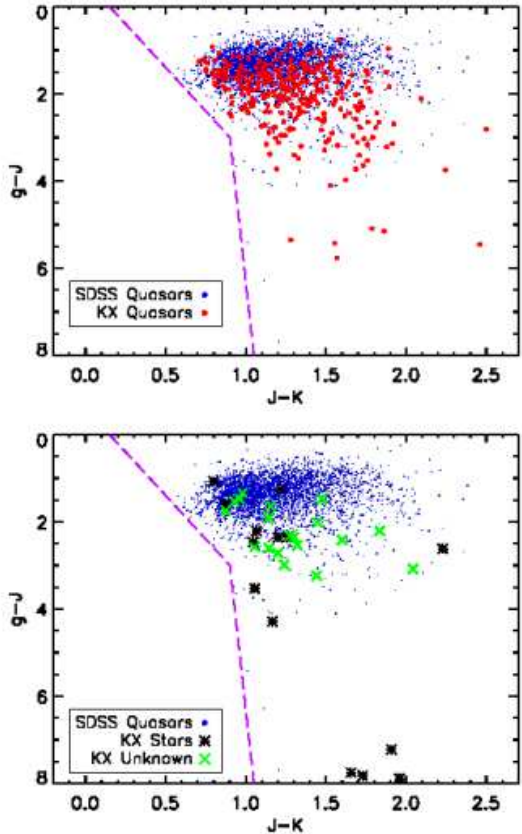}}
\caption{$gJK$ plots showing the location of the subgroups of
  objects. Top panel: the KX-only quasars (red dots) are redder than
  the SDSS quasars (blue dots). Bottom panel: the confirmed stars
  (black asterisks) are not confined to one section of the
  colour-space. The objects that have no spectral classification
  (green crosses) are located in the region of colour-space where
  highly reddened quasars overlap with low redshift galaxies.}
\label{fig:kx_gJK}
\end{figure}

The reduced, flux calibrated and Galactic extinction corrected spectra
are individually classified by hand, assigning each object an
identification of either quasar, emission line galaxy, absorption line
galaxy, star, or unclassified. The numbers of each class of object are
listed in Table~\ref{tab:specclass}, and the location of the quasars,
stars, and unclassified objects is shown in
Fig.~\ref{fig:kx_gJK}. Redshifts were assigned interactively by
displaying the spectrum and identifying emission or absorption
features. Table~\ref{tab:kxcat1} gives the column description for the
catalogue of new spectroscopically observed objects, including the
classification and a measure of the confidence of the
classification. The catalogue is available in the \underline{online
  version} of the paper.

The identification-redshifts are accurate to typically $\delta
z=\pm0.01$ and no correction has been made for the systematic
emission line shifts known to affect quasar redshift determinations
\citep{Hewett2010}. Such velocity shifts change the redshift estimates
by $\sim$\,10$^{-3}$, smaller than the uncertainty in the redshift
determinations from the visual identification process. If further
refinement in the redshift estimate is required, the catalogue records
the emission line from which the redshift was computed.

No distinction is made here between lower luminosity AGN and high
luminosity quasars, which for consistency with the SDSS quasar
catalogue has been set at M$_i \le -22.4$\footnote{Recall that all
  magnitudes quoted in this work are on the Vega system.}. In practice
all the confirmed broad-line objects have absolute magnitudes brighter
than M$_i = -23$ and are thus classified as quasars according to the
SDSS convention.

\begin{table}
\centering
  \caption{\label{tab:specclass}Spectroscopic classification of the 324
    objects for which new spectra were obtained}
\begin{tabular}{lrr} \hline
Class & Number & Percent \\ \hline
Quasars & 284 & 87.7 \\
Emission line galaxies & 2 & 0.6 \\
Absorption galaxies & 5 & 1.5 \\
Stars & 14 & 4.3 \\
Unclassified & 19 & 5.9 \\ \hline
Total & 324 & 100 \\ \hline
\end{tabular}
\end{table}

\begin{table*}
\centering
\caption{\label{tab:kxcat1}Catalogue of 324 KX-selected objects column format
  and descriptions}
\begin{tabular}{rcl}\hline
Column & Format & Description \\ \hline
1 & A8 & Unique KX Catalogue ID \\
2 & F12.6 & Right ascension in decimal degrees (J2000) \\
3 & F12.6 & Declination in decimal degrees (J2000) \\
4 & F9.6 & SDSS $u$-band magnitude, converted to Vega, corrected for
Galactic extinction \\
5 & F9.6 & SDSS $g$-band magnitude, converted to Vega, corrected for
Galactic extinction \\
6 & F9.6 & SDSS $r$-band magnitude, converted to Vega, corrected for
Galactic extinction \\
7 & F9.6 & SDSS $i$-band magnitude, converted to Vega, corrected for
Galactic extinction \\
8 & F9.6 & SDSS $z$-band magnitude, converted to Vega, corrected for
Galactic extinction \\
9 & F9.6 & UKIDSS $Y$-band magnitude, corrected for Galactic extinction \\
10 & F9.6 & UKIDSS $J$-band magnitude, corrected for Galactic extinction \\
11 & F9.6 & UKIDSS $H$-band magnitude, corrected for Galactic extinction \\
12 & F9.6 & UKIDSS $K$-band magnitude, corrected for Galactic extinction \\
13 & F9.6 & SDSS $u$-band magnitude error\\
14 & F9.6 & SDSS $g$-band magnitude error\\
15 & F9.6 & SDSS $r$-band magnitude error\\
16 & F9.6 & SDSS $i$-band magnitude error\\
17 & F9.6 & SDSS $z$-band magnitude error\\
18 & F9.6 & UKIDSS $Y$-band magnitude error\\
19 & F9.6 & UKIDSS $J$-band magnitude error\\
20 & F9.6 & UKIDSS $H$-band magnitude error\\
21 & F9.6 & UKIDSS $K$-band magnitude error\\
22 & I1 & Region flag \\
23 & F9.6 & Spectroscopic redshift \\
24 & F9.6 & NIR Photo-z Photometric redshift \\
25 & I1 & Classification \\
26 & I1 & Confidence of classification \\
27 & A7 & Identified Spectral Feature \\
28 & A20 & SDSS Object ID \\
29 & A12 & UKIDSS WSA Object ID \\
30 & A3 & Comment \\ \hline
\end{tabular}
\end{table*}

\subsection{Spectroscopic Completeness}

Of the 324 spectra, only 19 remain without identification. The spectra have no
classification primarily due to a lack of identifiable spectral
features rather than low signal-to-noise ratio (S/N).

The unclassified objects have similar distributions as the general KX object
population in their $i$- and $K$-band magnitudes, as well as their photometric
redshifts. The bottom panel of Fig.~\ref{fig:kx_gJK} shows the
location of the unclassified 
objects in the $gJK$ plane with respect to the other objects. The bluest
objects are in the same area as several quasars, but also some stars, and the
reddest objects are in the area of low redshift star-forming
galaxies. 

While definitive classifications are not possible for the 19 spectra,
it is possible to eliminate a range of potential identifications.  The
lack of evidence for a 4000\,\AA\ break in the spectra rules out
compact absorption line galaxies at $z < 0.7$, and at redshifts
greater than this the objects would be implausibly bright, given the
flux limits. Additionally, the H$\alpha$, \oiii 5008\,\AA\ and \oii
3728\,\AA\ emission lines would be visible in star-forming galaxies to
$z=$~0.1, 0.45 and 0.9, respectively. The absence of narrow emission
lines in the spectra, with the possible exception of two objects,
KXu\_0009 and KXu\_0013, disfavours identification as emission
line galaxies.

The absence of a Lyman-$\alpha$ break in the spectra indicate that all
the unidentified objects are at $z < 2$. Many of the spectra show
structure in their continua, which could be due to the presence of
\feii and \feiii emission in quasars, particularly in the redshift
interval $z\sim 1.0-1.5$, that possess relatively weak broad emission
lines. Such objects are not uncommon and a number of objects with
weak emission lines and/or unusual iron emission within the SDSS
spectroscopic database have recently been catalogued by
\citet{Meusinger2012}. 

Tentative `identifications' have been included in the comment column
in Table~\ref{tab:kxcat1} for the 19 objects: `?' (unknown); `g?' (galaxy?); `q?'
(quasar?); `q z?' (quasar with unknown redshift). More reliable
identifications will require spectra with higher S/N. 

The fraction of unidentified objects is small, just 6~per~cent, and
the statistics quoted for the success rate of the KX-selection and the
effective area of the survey are given assuming that none of the 19
unidentified spectra are quasars with $z>1.0$. In practice we estimate
that half the unidentified objects are quasars, in which case the
relevant statistics for the success rate of the KX survey will improve
slightly.

\section{KX-Selected Quasar Catalogue}\label{sec:KXcatalogue}

In addition to the newly acquired spectra, there exist already extensive
spectroscopic observations within the survey area. The SDSS, 2SLAQ, as
well as the Pilot KX Survey, overlap with the area covered by the new
observations. The individual candidate selection criteria that were
employed for each of the four spectroscopic surveys differ, but the
NIR Photo-$z$ selection algorithm is applied to all sources
uniformly. The final KX-selected quasar catalogue (or KX catalogue
for brevity) is composed of all objects selected as quasar candidates
by the NIR Photo-$z$ algorithm that are spectroscopically confirmed as
quasars by any of the above-mentioned four surveys. 

All quasars that satisfy the selection outlined in
Section~\ref{subsec:candsel} are hereinafter referred to as
`KX-selected quasars', while objects that are identified as quasar
candidates only by the NIR Photo-$z$ and are subsequently confirmed as
quasars are referred to as `KX-only quasars'.

\subsection{KX-Only Quasars}\label{subsec:newkx}

As listed in Table~\ref{tab:specclass}, 284 new KX-only quasars were
identified. However, as seen in Fig.~\ref{fig:ivsK}, many of the
quasars are fainter than the completeness magnitude limits. There were
also a number of quasars for which, although they possessed a
$z_{phot}\ge1.0$, their spectroscopic redshift is $z_{spec}<
1.0$. These objects will be excluded from the statistically complete
sample.

\begin{figure}
\resizebox{\hsize}{!}{\includegraphics{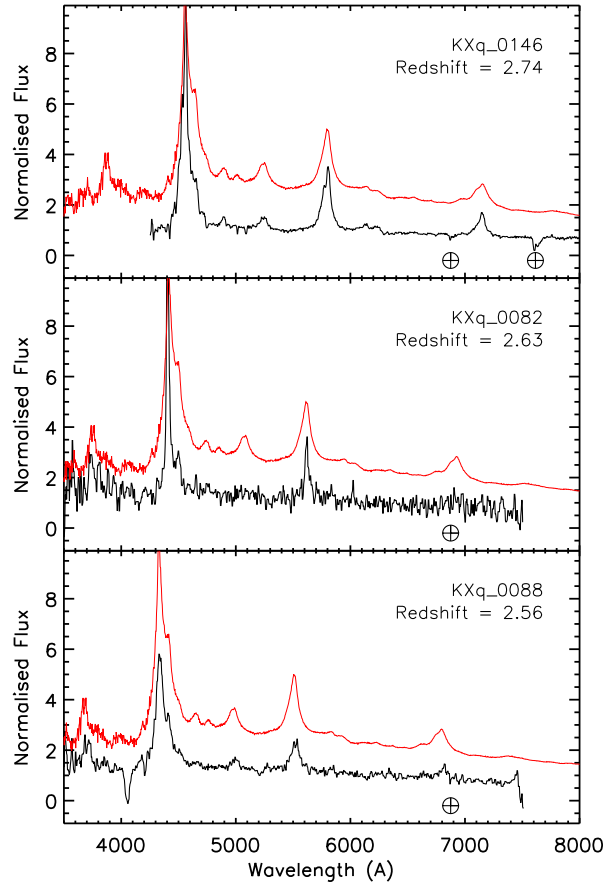}}
\caption{Example spectra of KX-only quasars. (Top) An otherwise
  ordinary quasar at $z\sim$ 2.7, the redshift at which optical
  selection suffers high incompleteness due to stellar contamination,
  (centre) a quasar with narrow emission lines, (bottom) a quasar with
  a strong intervening absorption system seen at 4050\,\AA. The SDSS
  quasar composite spectrum from \citet{VandenBerk2001} has been
  overplotted in red and offset for comparison.  The spectra are flux
  calibrated to flux per unit wavelength, and have been normalised and
  median smoothed for display. The catalogue object names and
  redshifts are given in each frame. The top spectrum is from the VLT
  FORS instrument, while the bottom two spectra are from the EFOSC
  instrument on the NTT. The locations of the telluric A and B bands
  are marked with the Earth symbols.}
\label{fig:spec1}
\end{figure}

\begin{figure}
\resizebox{\hsize}{!}{\includegraphics{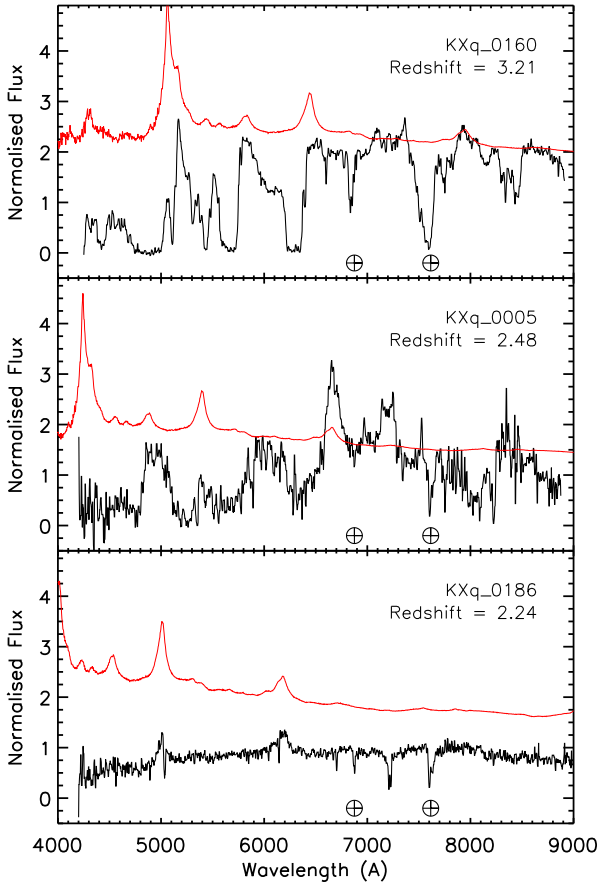}}
\caption{Example spectra of unusual KX-only quasars. (Top) A BALQSO,
  (centre) an extreme BALQSO, (bottom) a quasar with red spectral
  shape and strong intervening metal absorption lines. As in
  Fig.~\ref{fig:spec1}, the SDSS quasar composite spectrum has been
  overplotted in red for comparison. The spectra are flux calibrated
  to flux per unit wavelength, and have been normalised and median
  smoothed for display. The catalogue object names and redshifts are
  given in each frame. The top two spectra are from the VLT FORS
  instrument, while the bottom spectrum is from the MOSCA instrument
  on the Calar Alto 3.5m telescope. The locations of the telluric A
  and B bands are marked with the Earth symbols.}
\label{fig:spec2}
\end{figure}

Figs.~\ref{fig:spec1} and \ref{fig:spec2} show spectra for a selection
of the KX-only quasars. All of the subclasses of quasars that were
expected in the KX sample were indeed found.  Fig.~\ref{fig:spec1}
shows three quasars with redshifts $z\sim$\,2.7, placing them in the
region where the SDSS quasar-selection is known to be significantly
incomplete, as described in the Introduction. It was expected that a
large number of objects were found in this redshift range. The quasar
in the bottom panel is an example of a KX-only quasar with a strong
intervening absorption system.

The objects in Fig.~\ref{fig:spec2} show some of the more unusual
quasars uncovered by the KX selection. The top panel shows a strong
BALQSO, and the middle panel shows an even more extreme object, whose
redshift is somewhat uncertain. The bottom panel shows a quasar with
broad emission lines but a clearly curved continuum, likely indicating
a moderate amount of dust reddening.

\subsection{SDSS Quasars}\label{subsec:addsdss}

The existing SDSS quasars contribute the most objects to the final
combined KX catalogue. There are 3267 spectroscopically confirmed
quasars in the survey area, of which 1135 satisfy the appropriate $i$-
and $K$-band magnitude limits for the three survey areas and $1.0\le
z\le3.5$. The method with which the objects were selected by the SDSS
is not important, as all of the objects included in the KX catalogue
have also been selected by the NIR Photo-$z$ algorithm.

\subsection{Other Quasars in the Area}\label{subsec:addothers}

The 2SLAQ survey covers a two-degree wide strip of sky around
declination zero, with optical magnitude limits fainter than those of
the main SDSS quasar survey. 33 2SLAQ quasars were chosen by the NIR
Photo-$z$ selection within the survey area, but only one satisfies the
NIR Photo-$z$ selection criteria to the magnitude and redshift limits
imposed.

The data from the Pilot KX Survey also covers a strip around
declination zero, along with a small area of coverage at declination
$\delta=6$. An extra 35 quasars are added to the new KX catalogue,
with three satisfying the completeness limits. Table~\ref{tab:allspec}
summarises the number of quasars from the different surveys
contributing to the final KX-selected quasar catalogue.

\subsection{SDSS \texttt{TARGET} Flags}\label{subsec:TARGETs}

Many more objects within the SDSS photometric database were flagged as
quasar candidates by the SDSS quasar selection algorithm than could be
observed within the SDSS survey. Therefore, there is a population of
objects for which an SDSS quasar target flag is set but no SDSS
spectrum exists, hereinafter referred to as SDSS \texttt{TARGETs}. As
the objects chosen by the tiling algorithm for spectroscopic
observations from the list of targets have no special properties with
respect to the general pool from which they are drawn, it is expected
that, statistically, the objects will have the same properties as the
sample with spectra.

The efficiency of each of the SDSS \texttt{TARGET} flags can be
investigated by comparing all objects with the flag set to the
spectroscopic identification of the subset of objects that were
observed. It was found that some of the flags are more reliable than
others. The quasar target flags of \texttt{TARGET\_QSO\_CAP},
\texttt{TARGET\_QSO\_FAINT}, \texttt{TARGET\_QSO\_FIRST\_CAP},
\texttt{TARGET\_QSO\_HIZ}, \texttt{TARGET\_SERENDIP\_BLUE},
\texttt{TARGET\_SERENDIP\_FIRST}, and \texttt{TARGET\_ROSAT} were all
found to be reliable indicators of the target being confirmed as a
quasar, when the objects were also selected as quasar candidates by
the NIR Photo-$z$ selection.

Without exception, when the flag \texttt{TARGET\_QSO\_REJECT} was set
for an object, the probability that the object was confirmed as a
quasar was very low, i.e. the efficiency of this flag was only a few
per~cent. This is true even if the object was additionally selected as
a quasar candidate by the NIR Photo-$z$. The finding is entirely in
accord with expectation \citet{Richards2002} and objects with the
\texttt{TARGET\_QSO\_REJECT} flag set were removed from
consideration. Only 6~per~cent of the SDSS \texttt{TARGETs} were
removed for this reason.  Additionally, candidates with either
\texttt{TARGET\_REDDEN\_STD}, \texttt{TARGET\_STAR\_BHB}, or
\texttt{TARGET\_STAR\_CARBON} set, along with none of the quasar
target flags, were removed. These further reduce the number of targets
by 3~per~cent.

The efficiency of the NIR Photo-$z$ selection with respect to known
SDSS quasars, computed in Section~\ref{subsec:efficiency}, is very
high, at $\sim$\,99~per~cent. It is then reasonable to expect that the
efficiency with which the NIR Photo-$z$ correctly selects quasars from
the SDSS \texttt{TARGETs} will be equally high. Thus the number of
additional `quasars' that can be added into the catalogue is
\textit{selection efficiency} $\times$ N(\texttt{TARGETs}). Using the
efficiency of 99~per~cent, and finding that there are 1210 SDSS
\texttt{TARGETs} in the survey area, an extra $\sim$\,1198 `quasars'
are added to the catalogue. All 1210 \texttt{TARGETs} are retained in
the final KX quasar catalogue for completeness.

Although objects that possess SDSS quasar \texttt{TARGET} flags were,
in general, excluded from further spectroscopic observation, 28 SDSS
\texttt{TARGETs} were deliberately observed. Twenty-seven of the
targets are classified as quasars, for an efficiency of 96~per~cent,
consistent with the 99~per~cent computed from the SDSS
spectroscopically observed population. These 27 quasars reduce the
number of purely KX-selected quasars from 284 to 257, and have been
added to the SDSS spectroscopically confirmed quasars to increase
their number from 3267 to 3294, as noted in Table~\ref{tab:allspec}.

The statistical properties of the sample of objects with
identifications by NIR Photo-$z$ as quasars and with SDSS quasar
\texttt{TARGET}, are expected to be identical, irrespective of whether
objects possess an SDSS spectrum.

\subsection{The Final KX-Selected Quasar Catalogue}

The final KX-selected quasar catalogue composed of quasars selected by
the NIR Photo-$z$ routine, but spectroscopically observed within the
SDSS, 2SLAQ, Pilot KX Survey and the present large area KX survey
contains 3619 spectroscopically confirmed quasars, or 4829 quasars
with the SDSS \texttt{TARGETs} included. Of these, 1328 quasars have
spectra confirming they are at $1.0\le z\le3.5$, $K\le16.5$ and
$i\le19.5$ if they fall within the CA area, $K\le16.5$ and $i\le19.7$
if they are in the NTT/VLT area, and $K\le16.6$ and $i\le22.0$ within
the NTT/VLT Deep area.

Table~\ref{tab:kxcat2} gives a description of the columns contained
within the KX-selected quasar catalogue. As with
Table~\ref{tab:kxcat1}, the catalogue is available in the \underline{online
  version} of the paper. Additional information for each object can be
extracted from the SDSS and UKIDSS databases using the unique object
identifiers provided. Note that the UKIDSS DR4 database should be
queried when using the UKIDSS WSA Object IDs, as they are specific to
each data release.

\begin{table}
\centering
  \caption{\label{tab:allspec}Spectroscopically confirmed quasars and
    SDSS \texttt{TARGETs} contributing to the final KX-selected quasar
    catalogue. The numbers in parentheses account for the 27 KX
    quasars that also have SDSS \texttt{TARGET} flags set
    (Section~\ref{subsec:TARGETs}).  The third column indicates the
    number of objects from each survey that satisfy the magnitude
    limits for the CA, NTT/VLT or NTT/VLT Deep areas.}
\begin{tabular}{lrr} \hline
Survey & Total & Complete \\ \hline
Large Area KX & 284 (257) & 172 \\ 
SDSS & 3267 (3294) & 1152 \\
2SLAQ & 33 & 1 \\
Pilot KX & 35 & 3 \\ \hline
Subtotal & 3619 & 1328 \\ \hline
\texttt{TARGETs} & 1210 & 163 \\ \hline
Total & 4829 & 1491 \\ \hline
\end{tabular}
\end{table}

\begin{table*}
\centering
\caption{\label{tab:kxcat2}KX-selected quasar catalogue column format
  and descriptions}
\begin{tabular}{rcl}\hline
Column & Format & Description \\ \hline
1 & A8 & Unique KX Catalogue ID \\
2 & F12.6 & Right ascension in decimal degrees (J2000) \\
3 & F12.6 & Declination in decimal degrees (J2000) \\
4 & F9.6 & SDSS $u$-band magnitude, converted to Vega, corrected for
Galactic extinction \\
5 & F9.6 & SDSS $g$-band magnitude, converted to Vega, corrected for
Galactic extinction \\
6 & F9.6 & SDSS $r$-band magnitude, converted to Vega, corrected for
Galactic extinction \\
7 & F9.6 & SDSS $i$-band magnitude, converted to Vega, corrected for
Galactic extinction \\
8 & F9.6 & SDSS $z$-band magnitude, converted to Vega, corrected for
Galactic extinction \\
9 & F9.6 & UKIDSS $Y$-band magnitude, corrected for Galactic extinction \\
10 & F9.6 & UKIDSS $J$-band magnitude, corrected for Galactic extinction \\
11 & F9.6 & UKIDSS $H$-band magnitude, corrected for Galactic extinction \\
12 & F9.6 & UKIDSS $K$-band magnitude, corrected for Galactic extinction \\
13 & F9.6 & SDSS $u$-band magnitude error\\
14 & F9.6 & SDSS $g$-band magnitude error\\
15 & F9.6 & SDSS $r$-band magnitude error\\
16 & F9.6 & SDSS $i$-band magnitude error\\
17 & F9.6 & SDSS $z$-band magnitude error\\
18 & F9.6 & UKIDSS $Y$-band magnitude error\\
19 & F9.6 & UKIDSS $J$-band magnitude error\\
20 & F9.6 & UKIDSS $H$-band magnitude error\\
21 & F9.6 & UKIDSS $K$-band magnitude error\\
22 & I1 & Region flag \\
23 & I1 & Source of classification \\
24 & F9.6 & Spectroscopic redshift \\
25 & F9.6 & NIR Photo-z Photometric redshift \\
26 & I1 & Completeness flag \\
27 & F9.6 & $E(B-V)$ estimate \\
28 & I1 & BALQSO flag \\
29 & A20 & SDSS Object ID \\
30 & A12 & UKIDSS WSA Object ID \\
31 & A20 & SDSS SpecObjID \\ \hline
\end{tabular}
\end{table*}


\section{Evaluating the NIR Photo-$z$ Algorithm}\label{sec:photozeval}

The two quantities used to determine the overall quality of a
selection scheme are the {\it completeness} and the {\it
  efficiency}. The completeness describes the number of quasars that
are selected compared to the total number that exist but do not meet
the selection criteria, whereas the efficiency compares the number of
quasars identified with respect to the total number of candidates
chosen. Clearly, the aim is for both values to be as close to
100~per~cent as possible, but generally a compromise must be made,
with higher completeness achieved at the expense of lower efficiency,
and \textit{vice versa}.

\subsection{Recovery of SDSS Quasars}\label{subsec:completeness}

Determining the completeness of a selection algorithm is difficult as
the true number of objects at the appropriate magnitudes and redshifts
is required. Extensive analysis of the completeness of the SDSS quasar
selection routine has been carried out by \citet{Richards2006a},
involving a grid of simulated quasars covering a range of photometric
properties. However, the simulations do not include dust extinction,
nor do they allow for curvature in the quasar SED or BAL troughs, and
are thus only upper limits on the completeness.

A completeness calculation similar to that of \citet{Richards2006a} is
beyond the scope of the current work. However, it is still important
to confirm that the NIR Photo-$z$ routine is able to recover known
quasars as well as finding additional quasars. The SDSS spectroscopic
dataset is useful for this task due to the large number of objects and
the broad range of quasar SEDs included.  That said, it is important
to remember that the goal of the present study is to identify quasars
{\it not} contained within the SDSS catalogue.

The completeness of the NIR Photo-$z$ code with respect to the known
SDSS quasars is determined by crossmatching the quasar list to the UKIDSS LAS
DR4 database to extract their $YJHK$ photometry and morphology. The
NIR Photo-$z$ algorithm was then applied to the SDSS-UKIDSS photometry
for the SDSS spectroscopic sample. 

Within the survey area, 96.3~per~cent of the SDSS quasars with $0<z<5$
and a corresponding entry in the UKIDSS LAS are correctly identified
as a quasar candidate by the NIR Photo-$z$ selection. Restricting the
test to $1.0\le z\le3.5$, $K\le16.7$ and $i\le22.0$ and requiring that
the object is rightward of the selection line on the $gJK$ plot, which
more accurately reflects the final selection criteria of the
candidates, the completeness with respect to known SDSS quasars is
found to be extremely high, at 97.9~per~cent.

\begin{figure}
\resizebox{\hsize}{!}{\includegraphics{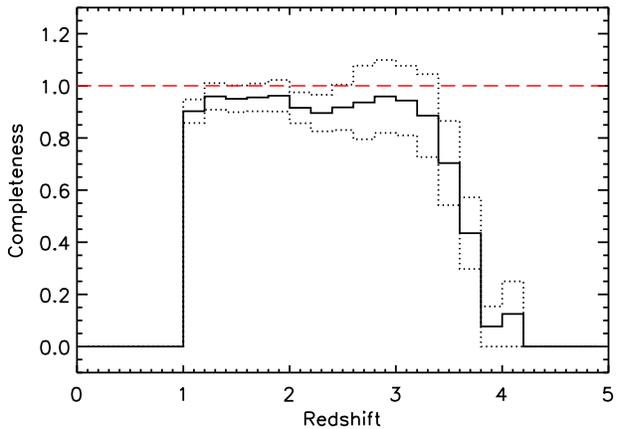}}
\caption{The completeness of the NIR Photo-$z$ quasar selection
  routine as determined by the successful recovery of known SDSS
  quasars, requiring that both $z_{spec}\ge1.0$ and
  $z_{phot}\ge1.0$. The 1\,$\sigma$ Poisson counting errors are shown
  as dotted lines. The completeness remains fairly constant until it
  drops off sharply at $z>3.5$.}
\label{fig:completeness2}
\end{figure}

A $\simeq$\,98~per~cent success rate is encouraging but the figure does
not exactly match the full selection criteria, which include the
restriction that the the photometric redshift, $z_{phot}$, must also
exceed unity. Requiring that the recovered quasars have $z_{phot}\ge
1.0$ reduces the completeness to 92.0~per~cent.
Fig.~\ref{fig:completeness2} shows the roughly constant selection
completeness to $z\sim$ 3.5, where the completeness drops off
drastically due partly to employing the $g$-band in the selection
(within the $gJK$ diagram) and a sub-optimal scheme for incorporating
non-detections in the $u$- and $g$-bands in the Photo-$z$ algorithm.

\subsection{Selection Efficiency}\label{subsec:efficiency}

In addition to the completeness, the other important quantity to
determine for the survey is the efficiency of the selection, which
relates the number of candidates chosen to the number of those
candidates that are confirmed as quasars. Using the extensive
spectroscopic information already available from the SDSS, 2SLAQ and
Pilot KX surveys in addition to the new spectroscopic observations, an
estimate of the overall efficiency of the selection algorithm can be
made.

There exist 3762 objects within the survey area that are selected as
quasar candidates by the NIR Photo-$z$ algorithm that also possess
spectra from one of the four above mentioned surveys, of which 3619
are confirmed as quasars. This results in a selection efficiency of
96.2~per~cent. Restricting the sample to the magnitude limits
appropriate for each area and requiring the candidate to have
$z_{phot}\ge1.0$, 97.4~per~cent of the candidates are confirmed as
quasars. The efficiency in the NTT/Deep region does not drop below
that of the other shallower regions. If a further restriction of
$z_{spec}\ge1.0$ on the confirmed quasar is applied, the efficiency is
reduced to 84.3~per~cent, due to the NIR Photo-$z$ correctly
identifying the quasar but incorrectly assigning a high
($z_{phot}>1.0$) redshift when the spectroscopic redshift is $z_{spec}<1.0$. 

Restricting the efficiency calculation to the new KX-only quasars, 284
of 324 candidates are confirmed quasars, for 87.6~per~cent
efficiency. For objects with $i\le19.5$, $K\le16.5$ and
$z_{spec}\ge1.0$, the efficiency drops to 75~per~cent, however, the
figure is encouraging given all such objects eluded the SDSS quasar 
selection algorithm. These values are lower limits, as there are a
number of probable quasars among the 19 objects that have no secure identification.

The KX-selected quasar candidates with new spectra classified as
stars, galaxies or unclassified (40 of 324 objects observed), do not
show clustering in any of their relevant properties such as $i$- or
$K$-band magnitudes, $z_{phot}$, or the NIR Photo-$z$ best matched
quasar model.  Further investigation is required to determine why
these objects were selected as quasar candidates, and how the number
of non-quasars can be reduced, thereby increasing the efficiency.

The selection efficiency does not show any strong dependence on
$i$-band magnitude, $z_{phot}$ or $i-K$ colour within the ranges
probed by the observations. This is encouraging for future efforts
that aim specifically at faint or very red objects.

\subsection{NIR Photo-$z$ Redshift Accuracy}\label{subsec:sdssphotoz}

It has been shown that the NIR Photo-$z$ routine is very good at
selecting known quasars as quasar candidates, i.e. the {\it
  identification} of the objects is reliable. Now we need to determine
the accuracy of the photometric redshifts, $z_{phot}$, of the quasar
candidates. This is particularly important because of the $z_{phot}\ge
1.0$ restriction imposed on the quasar candidate list.

\begin{figure}
\resizebox{\hsize}{!}{\includegraphics{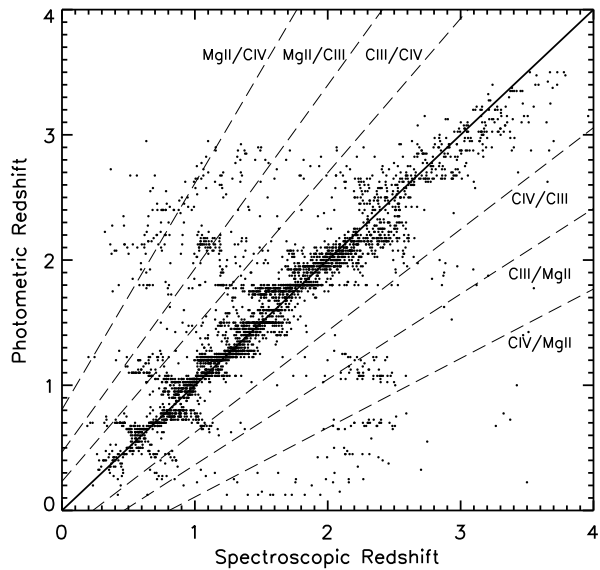}}
\caption{Comparison of the photometric redshifts compared to the spectroscopic
  redshifts for the confirmed quasars that are also chosen as quasar
  candidates by the photometric redshift routine. The solid diagonal line
  marks the one-to-one relation, while the dashed lines show the redshift
  relations for misidentified emission lines. The first emission line
  listed of the pair is the correct line identification, and the second line is
  the incorrect photometric redshift identification,
  i.e. spectroscopic ID/photometric ID. The horizontal stripes seen are
  a result of the $\Delta z=0.025$ sampling of the models in redshift.}
\label{fig:zcomparison1}
\end{figure}

\begin{figure}
\resizebox{\hsize}{!}{\includegraphics{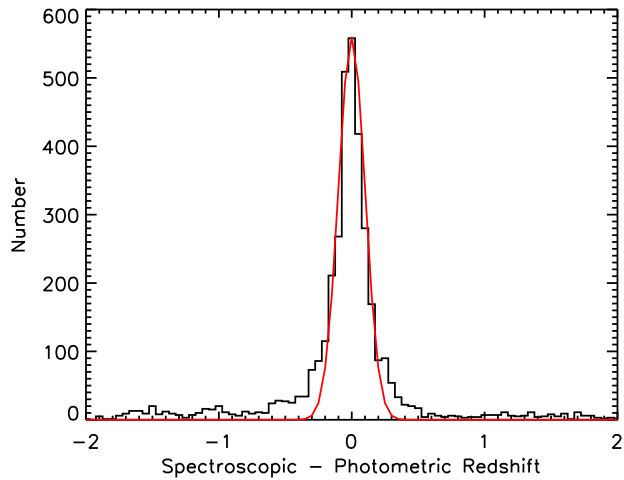}}
\caption{Histogram of the difference between the spectroscopic and
  photometric redshifts for the confirmed quasars shown in
  Fig.~\ref{fig:zcomparison1}. The distribution is centred at zero and
  the core has a width of $\sigma=0.1$, shown by the Gaussian
  distribution overplotted in red. The few objects with large positive
  values would be excluded by the $z_{phot}\ge1.0$ restriction.}
\label{fig:zcomparison2}
\end{figure}

For all of the 3619 quasars in the final KX catalogue, the $z_{phot}$
and $z_{spec}$ can be compared, as shown in
Fig.~\ref{fig:zcomparison1}. Overall, the NIR Photo-$z$ code is fairly
accurate at determining the redshifts of known quasars; the quasars
follow the the $z_{spec} = z_{phot}$ line with no significant
systematic offset. The core of the distribution is well fit by a
Gaussian centred at zero with $\sigma=0.1$, as shown in
Fig.~\ref{fig:zcomparison2}. The small-scale wiggles as a function of
redshift can be attributed to emission lines moving into and out of
the various passbands. There are, however, a fraction of outliers,
referred to as `catastrophic failures'; 20.6, 14.1 and 8.6\,per~cent
of the quasars in Figs.~\ref{fig:zcomparison1} and
\ref{fig:zcomparison2} have spectroscopic redshifts that differ from
the photometric redshift by more than $\Delta z=$ 0.3, 0.5 and 1.0,
respectively. The small clusters of objects result from confusion
caused by the misidentification of emission lines, indicated on the
plot by the dashed lines.

The redshift accuracy of the NIR Photo-$z$ appears to perform at least
as well, if not better, as the Bayesian classification algorithm based
only on optical colours from the study of \citet{Richards2009}. The
widths of the $z_{spec} - z_{phot}$ distributions are comparable and
both centred around zero, but the fraction of objects with $\Delta z \ge 1.0$
using optical colours alone is 11.6~per~cent, larger than the
8.6~per~cent for the NIR Photo-$z$ routine. The difficulty in
selecting quasar candidates at $2.5<z<3.0$ using only optical colours 
is also evident in the quasar candidate catalogue constructed by
\citet{Richards2009}, whereas the NIR Photo-$z$ does not experience
such redshift-dependent efficiency or completeness. These
results reinforce the conclusions drawn from the study of
\citet{Hatziminaoglou2000}, where the combination of NIR and optical
bands provides a clear advantage over using optical photometry alone.

\subsection{Photometric Variability}\label{subsec:variability}

As quasars are known to vary in brightess with time, and the optical
and NIR photometry used within the current study are not taken
simultaneously, the effects of this variability should be
investigated. Using magnitude measurements from two epochs for 25\,000
SDSS quasars, \citet{VandenBerk2004} have determined that quasar
variability amplitude is smaller at longer wavelengths, and
variability amplitude is smaller for brighter objects. Both of these
trends are beneficial to the current study.  As shown in the previous
sections, both the completeness and efficiency of the selection are
extremely high, so changes to intrinsic optical-NIR colours due to 
photometric variability must be sufficiently small such that the
selection of quasars is largely unaffected.

The requirement a candidate lies rightward of the selection boundary
in the $gJK$ plot could also potentially exclude a small number of
objects due to variability. Empirically, the fraction of known SDSS
quasars falling leftward of the selection boundary is less than
1~per~cent (Section~\ref{subsec:candsel}) strongly suggesting that the
effect of variability is minimal. The very small loss of quasars is
consistent with the locataion of quasars within the $gJK$-space.
Horizontal movement in the plot is due to a change in $J-K$ colour,
which is expected to be very small. Vertical movement due to varying
$g-J$ colour is also expected to be small, $\sim$\,0.1 magnitude, for
the rest-frame time intervals for these objects. Only at $g-J<3$ would
vertical movement on the $gJK$ plot cause an object to cross the
selection line. Objects lying close to the selection line at $g-J<3$
are blue and have $z<1$, and are therefore very likely to have been
selected by the SDSS already.

\section{Properties of KX Quasars}\label{sec:kxproperties}

The primary rationale for the current work is to present a description
of the selection of the KX quasar sample along with the catalogue of
objects resulting from the spectroscopic investigation of the
sample. However, below we outline some general properties of the
KX-only quasars with respect to optically selected SDSS quasars.

\subsection{Redshift Distribution}

\begin{figure}
\resizebox{\hsize}{!}{\includegraphics{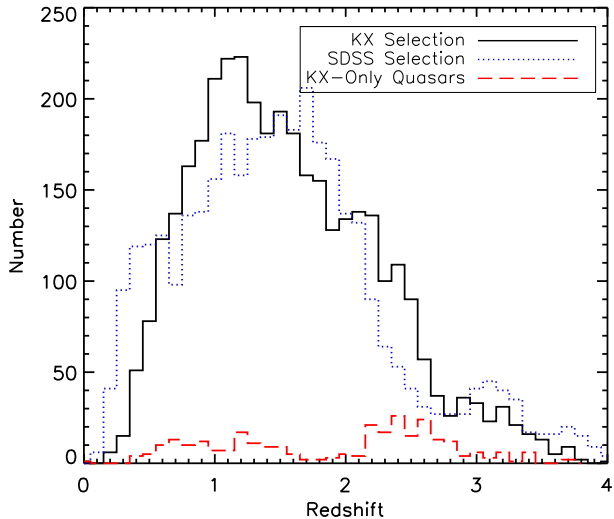}}
\caption{Redshift histogram of the KX-selected quasar catalogue,
  including both the new KX-only quasars and SDSS quasars selected by
  the NIR Photo-$z$ routine (black solid line), the SDSS quasars from
  the entire SDSS DR7 quasar catalogue (blue dotted line) scaled down
  by a factor of 30 for plotting, and the new
  KX-only quasars (red dashed line).}
\label{fig:zhist1}
\end{figure}

Fig.~\ref{fig:zhist1} shows three redshift histograms for different
quasar samples. The black solid line shows the redshift distribution
of all the quasars in the area including new KX and SDSS objects that
have been uniformly selected by the NIR Photo-$z$ algorithm. The blue
dotted histogram is for the enhanced SDSS DR7 quasar catalogue from
\citet{Hewett2010}, scaled down by a factor of 30 for plotting. The
well-known deficits of objects at $z\sim$ 2.7 and $z\sim$ 3.5 are
clearly visible in the SDSS histogram. The red dashed histogram shows the new
KX-only quasars. The redshift distribution of the KX-only plus SDSS
NIR Photo-$z$ selected quasars is smoother than that of the
SDSS-selected quasars, with the deficit of objects at $z\sim$ 2.7
greatly reduced once the new KX-selected quasars are incorporated.

\begin{figure}
\resizebox{\hsize}{!}{\includegraphics{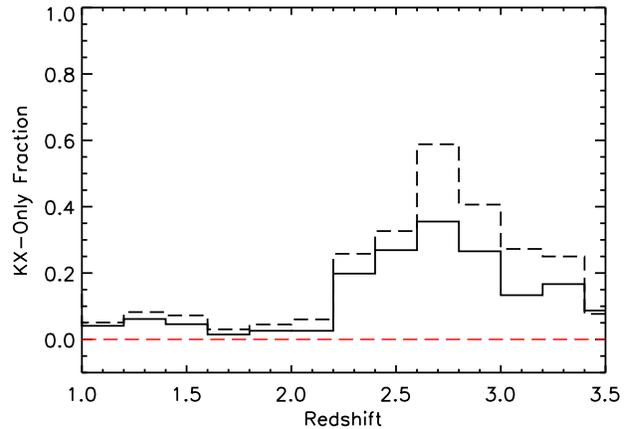}}
\caption{Fraction of the full KX catalogue that are KX-only quasars
  (solid histogram). The dashed histogram shows the fraction when the
  catalogue is restricted to $i\le19.5$, $K\le16.5$ and $z_{spec}$ and
  $z_{phot}\ge1.0$. The fraction of KX-only quasars reaches as high as
  63~per~cent for the redshifts at which the optical colours of
  quasars are similar to those of Galactic stars.}
\label{fig:zfrac}
\end{figure}

Fig.~\ref{fig:zfrac} shows the fraction of KX-only quasars
contributing to the KX catalogue at each redshift. As expected, the
KX-only quasars contribute the most at $2.5<z<3.0$ where the SDSS has
difficulty selecting objects. However, the excess of objects at
$z<1.6$ is also interesting, as the completeness of the SDSS selection
algorithm is considered to be very high at these redshifts. This is
discussed further in Section~\ref{subsec:extrakx}.

\subsection{$E(B-V)$ Distribution}

The $E(B-V)$ of each quasar can be estimated in a number of ways. For
this brief study, we choose to use the available photometry to
determine the amount of reddening present instead of, for example,
fitting a quasar model + dust extinction curve to the spectra. In
addition to being independent of the quality of the spectral flux
calibration, the photometry offers a much longer wavelength baseline
than the optical spectra, which is useful for constraining the amount
of dust present.

\begin{figure}
\resizebox{\hsize}{!}{\includegraphics{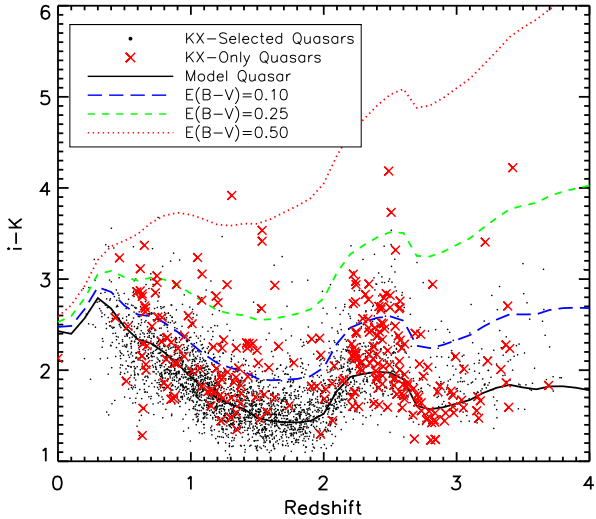}}
\caption{$i-K$ \textit{vs} redshift for the quasars in the KX
  catalogue.  The solid black line indicates the $i-K$ colour for a
  model unobscured quasar, the blue long dashed line indicates the
  same model quasar reddened by SMC-like dust with $E(B-V)=0.10$, the green
  short dashed line is the model quasar reddened with
  $E(B-V)=0.25$, and the red dotted line is the model quasar with
  $E(B-V)=0.50$. The majority of the quasars cluster around the
  unreddened model, as expected, whereas the KX-only quasars show a
  much broader range of $i-K$ colours.}
\label{fig:imkq}
\end{figure}

We use the $i-K$ colours of the quasars as a measure of the overall
colour of the quasars as it provides the longest baseline in
wavelength without being affected by absorption in the Ly$\alpha$
forest at high redshifts. Fig.~\ref{fig:imkq} shows the distribution
of $i-K$ colours of the quasars in the KX catalogue as a function of
redshift. Overplotted are the $i-K$ colours derived from a subset of
the model quasars originally used in the NIR Photo-$z$. The quasars
can be seen to occupy a wide range of $i-K$ values at fixed redshift.

Once the influence of quasar host galaxy light on the NIR-colours has
become small, at redshifts $z\ga1.2$, the main quasar SED-related
change to the observed $i-K$ colours is due to the presence of the
H$\alpha$ emission line in the $K$-band for the reshift interval
$2.1\la z \la 2.6$. The colour-change for a quasar with typical
H$\alpha$ emission-line strength is $\simeq$0.4\,mag. Emission-line
strength variations of up to a factor two are not uncommon and some
quasars can thus become redder by nearly a magnitude in $i-K$ due to
the presence of H$\alpha$ in the $K$-band. Fig.~\ref{fig:imkq} illustrates the
effect very clearly. 

It is important to note that two very different effects are
responsible for the excess of KX-only quasars with redshifts
$2.1<z<3.0$ visible in Figs.~\ref{fig:zhist1} and \ref{fig:zfrac}. The
first is the established incompleteness in the SDSS optical
colour-based selection which becomes particularly severe at 
redshifts $z>2.5$ (\citealt{Richards2002}). The second effect
results from the presence of the strong H$\alpha$ emission line in the
$K$-band, which defines the flux-limit of our sample. For a typical
quasar, the presence of the emission line results in an apparent
brightening within the redshift interval $2.1\le z\le2.6$. The $K$-band
sample thus reaches approximately 0.4\,mag deeper into the quasar
luminosity function, resulting in an apparent excess of KX-only
quasars within the interval $z=2.1-2.6$ of a factor $\simeq$2.

\begin{figure}
\resizebox{\hsize}{!}{\includegraphics{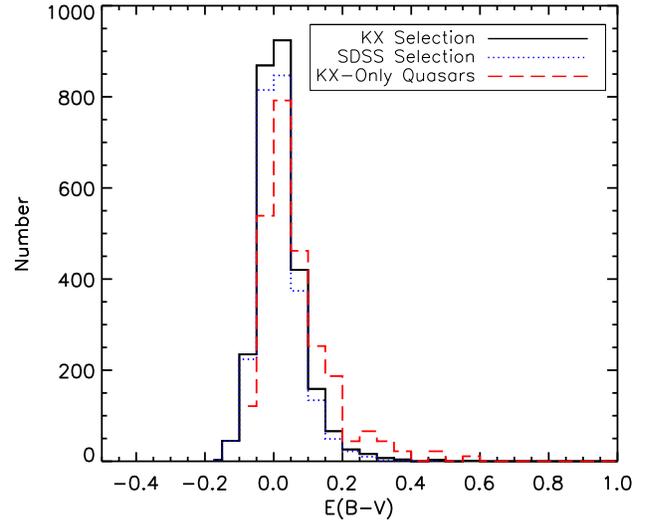}}
\caption{Histogram of $E(B-V)$ values derived from the $i-K$ colours
  of the $z\ge 1.0$ quasars shown in Fig.~\ref{fig:imkq} and as
  described in the text.  Negative values result from quasars with
  particularly blue continua. The KX-only quasars (red dashed
  histogram) are clearly redder than the SDSS quasars (blue dotted
  histogram), with a much more pronounced red tail extending to
  $E(B-V)\sim$ 0.4. The KX-only quasar histogram has been scaled up by
  a factor of 11 for display.}
\label{fig:ebvhist}
\end{figure}

From Fig.~\ref{fig:imkq}, an estimate of the amount of reddening each
object is experiencing can be made from its vertical location with
respect to the reddened quasar model loci. Fig.~\ref{fig:ebvhist}
shows the result of this procedure when applied to the KX quasars, the
SDSS quasars within the KX catalogue and the KX-only quasars at $z\ge
1.0$. The KX-only quasars are systematically redder than the
SDSS quasars, with a more pronounced red tail. The width of the
distribution in Fig.~\ref{fig:ebvhist} centred at $E(B-V)=0.0$
provides an estimate of the uncertainty in the $E(B-V)$ values
measured. The median $E(B-V)$ values for the SDSS quasars is 0.009,
whereas that for the KX-only quasars is 0.038. The histograms
significantly underestimate the fraction of reddened KX-only quasars
because, as is evident from Fig.~\ref{fig:imkq}, the large number of
`normal' quasars in the redshift interval $2.3<z<3.0$ missed by the
SDSS include many objects with essentially zero $E(B-V)$ estimates.

The estimates of $E(B-V)$ shown in Fig.~\ref{fig:ebvhist} account for
the presence and variation of strength of H$\alpha$ in the K-band at
$2.1\le z\le2.6$. For the majority of the  
objects within this redshift range, the standard quasar model, or a reddened
version of it, was the best fit within the NIR Photo-z routine. The
$E(B-V)$ values for these objects are computed with respect to the
standard and reddened standard quasar models. For the remaining
objects, the best fit was for a model with twice strong emission
lines, and their $E(B-V)$ values were derived with respect to this
model instead of the quasar model with standard emission line
strength. The stronger emission line models result in derived
$E(B-V)$ values that are smaller by between 0.02--0.05 compared to the
standard model.

Using photometry to determine $E(B-V)$ makes the assumption that any
significant deviation of the quasar $i-K$ colours from those expected
from the unreddened model arise from dust reddening. This assumption
is poor for low redshift ($z<1$) objects, as the quasar host galaxy flux can
boost the NIR flux, thus giving the impression of a redder SED that is
not representative of the SED shape of the quasar. The assumption also
does not account for the intrinsic spread of quasar spectral slopes at
optical wavelengths, which can be as large as $-1.0<\alpha_{\nu}<0.0$
for $F(\nu)\propto\nu^{\alpha}$ \citep{Richards2006b}.

Analysis of the $E(B-V)$ values derived from photometry requires all
of these effects to be considered. An accurate measure of the
reddening distribution of the KX-selected quasars is being undertaken
and the results will provide a quantitative determination of the
fraction of dust reddened quasars, with $E(B-V)\le0.5$\,mag, missing 
from optically selected samples.

\subsection{BALQSO Fraction}\label{subsec:balfrac}

The newest BALQSO catalogue available is that of \citet{Allen2011},
based on the SDSS DR6 quasar catalogue. This catalogue computes the
traditional \textit{balnicity index} (BI) for the Si\texttt{IV},
C\texttt{IV}, Al\texttt{III} and Mg\texttt{II} emission lines. In
a preliminary assessment of the observed frequency of BALQSOs we
focus only on absorption related to the C\texttt{IV} line here as
these troughs are easiest to identify and span a redshift range
similar to the objects in the new observations. It is assumed that
wavelength coverage to 1450\,\AA\ is required in order to identify a
C\texttt{IV} BAL trough in an SDSS spectrum, which corresponds to 
quasars at $z\ge1.62$. 

The NIR Photo-$z$ selected quasars with spectral coverage down to
1450\,\AA \ have been visually inspected to identify C\texttt{IV}
BALQSOs. No objects from the Pilot KX survey or 2SLAQ are included
here as there are only a few objects above $z=1.62$.

From the DR6 BALQSO catalogue, the computed mean \textit{observed}
BALQSO fraction
($f_{\textrm{BAL}}=N_{\textrm{BALQSO}}/N_{\textrm{total}}$) for
C\texttt{IV} BALQSOs over $1.5\le z\le 3.5$ is
$f_{\textrm{BAL}}=0.09$. Adding the visually identified KX BALQSOs to
the known SDSS BALQSOs, the observed mean BALQSO fraction within the
KX catalogue over the same redshift range increases to
$f_{\textrm{BAL}}=0.14$.

\begin{figure}
\resizebox{\hsize}{!}{\includegraphics{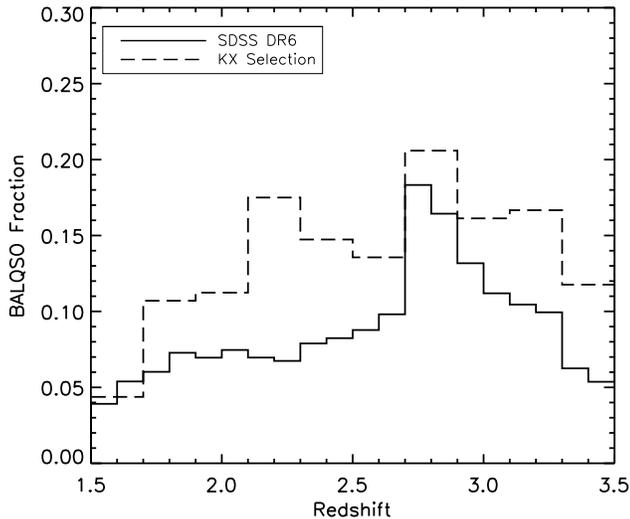}}
\caption{The observed fraction of BALQSOs as a function of redshift over the
  interval $1.5\le z \le 3.5$. The solid histogram is derived from the BALQSO
  catalogue from \citet{Allen2011}, the dashed histogram includes both SDSS
  and newly observed KX BALQSOs.}
\label{fig:balfrac}
\end{figure}

Fig.~\ref{fig:balfrac} shows the observed BALQSO fraction as a
function of redshift. The solid histogram is from the catalogue of
\citet{Allen2011}, whereas the dashed histogram includes known SDSS
and newly observed KX BALQSOs, both selected by NIR Photo-$z$. The bin
size for the KX+SDSS sample is $\Delta z=0.2$ to increase the number
of objects in each bin. The redshift bin from $1.5<z\le1.7$ is
incomplete, as SDSS C\texttt{IV} BALQSOs can only be reliably
identified at $z>1.6$.

The peak of BALQSOs in the SDSS histogram at $z\sim$ 2.7 is due to
the BAL troughs changing the optical colours of the objects enough to
remove them from the stellar locus. Thus at this redshift, BALQSOs are
preferentially selected while ordinary quasars are missed. Conversely,
the dip in the SDSS histogram at $z\sim$ 2.3 is due to the BAL
troughs altering the optical colours of the quasars such that they are
consistent with the stellar locus, so although ordinary quasars are
selected at this redshift, BALQSOs are preferentially missed by the
optical colour selection (see Fig. 21 in \citet{Allen2011} and the
associated discussion for an illustration of this effect). The new KX
quasars at $z\sim$ 2.3 are heavily dominated by BALQSOs. KX
selection of quasars is clearly a valuable tool for working toward an
accurate determination of the intrinsic fraction of BALQSOs.

\subsection{Dusty Absorbers}

There are 34 quasars with identified intervening absorbers, of which
10 have estimated $E(B-V) > 0.1$. As the new KX spectra are only of
moderate resolution, further observations at higher resolution and
higher S/N are required for analysis of the intervening absorbers.

\section{Discussion}\label{sec:discussion}

A number of extra quasars have been found by the NIR Photo-$z$
selection algorithm relying on the $K$-band excess of quasars with
respect to Galactic stars. However, as the completeness of the SDSS
quasar selection algorithm is known to be less than 100~per~cent, this
is unsurprising. Here we investigate the number of extra KX-only
quasars found with respect to the estimated SDSS completeness, and
consider why these objects were not optically selected as quasar
candidates.

\subsection{Extra KX Quasars}\label{subsec:extrakx}

The overall completeness of the multicolour and multiwavelength
selection employed by the SDSS (\citealt{Richards2002}) has been
computed to be 89~per~cent (\citealt{VandenBerk2005}). The
completeness is a strong function of redshift, with significantly
lower values for particularly problematic redshifts. The large
correction factors at problematic redshifts where the completeness can
be as low as 5~per~cent \citep{Richards2006a} make this number
uncertain.

It has been shown that KX selection is capable of identifying quasars
that were not flagged as quasar candidates by the SDSS targetting
routine. It is important to determine whether the number of extra
quasars is compatible with the expected completeness of the SDSS
selection, or whether a genuine unexpected excess of objects has been found.

\begin{figure}
\resizebox{\hsize}{!}{\includegraphics{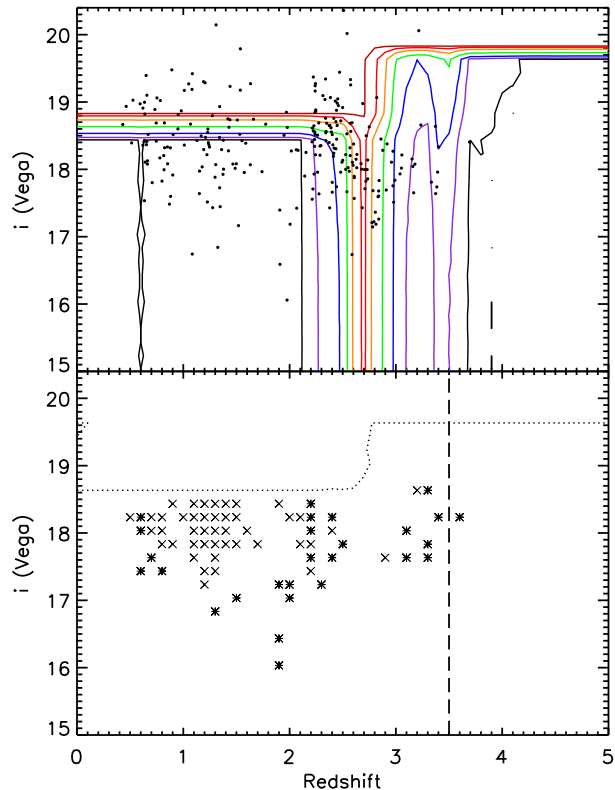}}
\caption{(Top) Completeness contours for point sources as computed in
  \citet{Richards2006a}, with the completeness line contours at 1
  (dark red), 10, 25, 50, 75, 90 and 99 (black) per cent. The location
  of the KX-only quasars in the $i-$redshift plane are
  overplotted. (Bottom) The excess of KX-only quasars with respect to
  the number of SDSS quasars expected to be missing based on the
  completeness, expressed as a percentage. Crosses indicate regions of
  the $i-$redshift plane where there is more than a 1~per~cent excess
  of KX-only quasars, and asterisks indicate the regions where there
  is more than a 10~per~cent excess. The black dotted line indicates
  the region of the plot where the SDSS completeness is zero, and for
  which this difference is not computed. The black dashed line at
  $z=3.5$ marks the upper redshift completeness limit of the new
  KX-only quasars.}
\label{fig:izcompl}
\end{figure}

Various estimates of the completeness of the SDSS target selection
algorithm have been performed (\citealt{Richards2002} and
\citealt{VandenBerk2005}, for example), but here we use the
completeness determination from the calculation of the SDSS quasar
luminosity function \citet{Richards2006a}. From that work, the
computed completeness between $15.0\le i\le 20.2$ and $0\le z\le6$ are
provided in tabular form for point sources, extended sources and
radio-selected sources. The point source completeness is used here, as
all the KX candidates are unresolved, and is shown in the top panel of
Fig.~\ref{fig:izcompl}. Overplotted is the location of the KX-only
quasars, which can be seen to primarily cluster around the problematic
redshifts of $2<z<3$, but there are still a significant number of
quasars at $z<2$ and $i<18.7$ where the SDSS completeness is expected
to be very high.

Within the area surveyed, for each $i$-band magnitude and redshift
grid point, the number of objects from the enhanced SDSS DR7 quasar
catalogue from \citet{Hewett2010} is determined, along with the
corresponding completeness. The grid cells are $\Delta i=0.2$ and
$\Delta z=0.1$ in size. Multiplying the number of SDSS quasars found
with the completeness provides the number of objects that are known to
be missing from the SDSS catalogue. For the same magnitude and
redshift cell, the number of KX quasars are also counted, and compared
with the known number of missed objects. The bottom panel of
Fig.~\ref{fig:izcompl} shows the magnitudes and redshifts for which
the number of KX-only quasars exceeds that expected from the
completeness. The black dotted line delimits the region of the plot
for which the SDSS completeness is zero.

As seen in the lower panel, the large number of KX-only quasars found
at $z\sim$ 2.7 are not beyond what is expected from the low SDSS
completeness. However, there is an excess at $z=2.3$, corresponding to
the redshift at which the new KX-only quasars are dominated by
BALQSOs. Quasars with BALQSO SEDs were not included in the SDSS
completeness corrections used here, and thus the completeness at this
redshift is underestimated. There is also an excess of KX-only quasars
at $z\sim$ 1. A number of these quasars have estimated reddening of
$E(B-V) > 0.1$ and their spectra show significant curvature. Objects
with curved SEDs were also not included in the completeness
simulations, so finding an excess of objects at this redshift is
unsurprising. However, it does show that there is a measurable
population of quasars at this redshift missing from the SDSS quasar
catalogue and the corresponding completeness calculations.

All of the quasars within the survey area from the enhanced SDSS DR7
quasar catalogue are used in the above calculation, including quasars
selected by serendipity, radio and X-ray properties, i.e. neither of
the two well-defined colour-selection algorithms. The completeness
values from \citet{Richards2006a} are for the colour-selection only
and the the lower panel of Fig.~\ref{fig:izcompl} 
underestimates the true number of excess KX quasars with respect to
the SDSS colour selection.

\subsection{SDSS Non-Selection}\label{subsec:nonSDSS}

Why were the KX-only quasars not selected as quasar targets by the
SDSS target algorithm? Broadly the explanations fit into three
categories. The first is that, at certain redshifts, the optical
colours of the objects place them in the stellar locus. The second is
that the objects are too faint in $i$-band to be selected. The third
is the most interesting case, i.e. the objects satisfy the SDSS
$i$-band magnitude limit(s) and do not lie within the stellar locus.

\begin{figure}
\resizebox{\hsize}{!}{\includegraphics{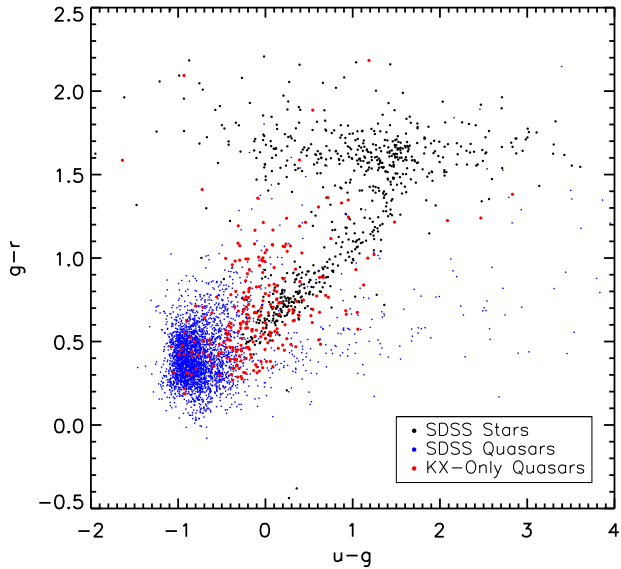}}
\caption{$u-g$ \textit{vs} $g-r$ colour-colour plot showing the location of
  SDSS quasars (blue dots) with respect to KX-only quasars (red dots) and SDSS
  stars, extracted from the SDSS database in order to show their
  location. The SDSS quasars lie in a distinct part of colour space
  from the stars. A large number, but not all, of the KX-only quasars overlap with the
  stellar locus.}
\label{fig:ugr}
\end{figure}

Fig.~\ref{fig:ugr} shows the $u-g$ \textit{vs} $g-r$ colour--colour
plot similar to that shown in \citet{Richards2002} displaying the
stellar locus (black dots) and the SDSS selected quasars (blue dots),
which generally do not overlap in colour-space. A large number of the
KX-only quasars (red dots) lie within the stellar locus, as
expected. However, the KX-only quasars located in the same region as
the SDSS quasars are potentially of considerable interest.

Extracting the KX-only quasars which lie in the same region of
colour-space as the SDSS quasars, the objects do not have particularly
large values of $E(B-V)$ or faint $i$-band magnitudes, i.e. most are
brighter than $i=18.7$ and not significantly reddened. Many of the
objects are BALQSOs at $2.2<z<2.7$ where the gradient in the SDSS
selection probability is relatively steep. It is likely that a small,
but systematic, error in the SDSS selection probability calculation is
responsible for the excess of quasars with somewhat atypical SEDs
found in the redshift interval $2.2<z<2.7$.

\subsection{Further Work}

Construction of the KX-selected quasar catalogue is an important first step
towards understanding the quasar population at $1.0\le z\le3.5$
with much reduced complicating factors of redshift-dependent selection
effects and objects being missed due to moderate amounts of dust
reddening. However, much more information can be derived not only from
the catalogue itself, but from further observations of specific
objects. Comparison of the spectral shape and emission line properties
of the KX-selected quasars with respect to
optically selected objects is currently underway. Perhaps information
from these further studies will help determine why the KX-only quasars were
not selected by the SDSS selection algorithm. 

The distribution of dust extinction experienced by the NIR-selected
quasars will allow a much better determination of the fraction of
quasars missed from current surveys due to reddening than is currently
possible with existing data. Further observations of the KX-only quasars
which have intervening absorption systems will also help determine
whether these absorbers are dustier than those found along the
line-of-sight to optically selected quasars. These topics are only a
few of the several interesting lines of research that can be
undertaken using the KX-selected quasar catalogue as a starting point.

Searching for quasars using combinations of optical and NIR colours is
a science goal for the Chinese Large Sky Area Multi-Object Fibre
Spectroscopic Telescope (LAMOST, \citealt{Su1998}). LAMOST is a purely
spectroscopic survey which must use existing imaging data to select
candidates for follow-up observations. Two primary sources of
photometric data for the survey are the SDSS and UKIDSS databases. Analysis
based on known SDSS quasars suggests using two different optical--NIR
colour-colour diagrams to select low and high redshift quasars, with
resulting combined completeness of 99.4~per~cent and contamination
from stars of 6.3~per~cent. However, these results only relate to
known SDSS quasars, and do not apply to objects that are not selected
as quasar candidates by SDSS. The NIR Photo-$z$ technique would be a
valuable tool for quasar selection for LAMOST.

\section{Conclusions}

Using a combination of optical and NIR photometry and a custom-built photometric
redshift routine to select quasar candidates at $z\ge 1$ has been shown
to be a valuable tool for collecting a quasar sample with fewer biases
than optical selection. Section~\ref{sec:photozeval} evaluated the NIR
Photo-$z$ routine and found it to be highly efficient at correctly
identifying candidates as quasars, with efficiency reaching 97~per~cent for
$z_{phot}\ge 1.0$, and 84~per~cent for objects with $z_{spec}$ and $z_{phot}\ge 1.0$. 
The selection algorithm is also complete with respect to known SDSS
quasars, correctly recovering 98~per~cent of quasars with $1.0\le z\le
3.5$, $K\le 16.7$ and $i\le 22.0$.

Interesting properties of the new KX-only quasars are outlined in
Section~\ref{sec:kxproperties}. A large number of extra quasars are
found at $z\sim$ 2.7 as expected as the SDSS selection is
particularly inefficient at this redshift, but excesses of KX-only
quasars are also found at redshifts where the SDSS completeness is
expected to be very high. Excesses of objects at $z=2.3$ and $z<1.6$
are found, due to the unusual spectral properties of KX-only quasars
at these redshifts.

The KX-only quasars also show indications of being redder than SDSS
quasars when comparing the $i-K$ colours of the two subgroups. Further
analysis of the reddening properties of the KX quasars will yield a
better estimate of the fraction of dust-reddened quasars currently
missing from optically selected surveys. The KX quasar selection is
also very effective in identifying BALQSOs, with virtually all of the
new quasars at $z\sim$ 2.3 showing BAL troughs. A number of quasars
with high equivalent width interving absorption systems have also been
found, some of which show signs of moderate ($E(B-V) > 0.1$)
reddening. Further observations at higher S/N and spectral resolution
are required to determine whether these absorbers contain significant
dust columns.

The KX-selected quasar catalogue, and NIR-selection of quasars, has
significant potential for constructing quasar catalogues less biased
by dust and other redshift-dependent selection effects that affect
optical quasar selection.

\section*{Acknowledgments}

We acknowledge the contributions of the staff of UKIRT to the
implementation UKIDSS survey and the Cambridge Astronomical Survey
Unit and the Wide Field Astronomy Unit in Edinburgh for processing the
UKIDSS data. This work is based in part on data obtained as part of
the UKIRT Infrared Deep Sky Survey. The United Kingdom Infrared
Telescope is operated by the Joint Astronomy Centre on behalf of the
UK Science and Technology Facilities Council. 

Funding for the SDSS and SDSS-II has been provided by the Alfred P. Sloan
Foundation, the Participating Institutions, the National Science Foundation,
the U.S. Department of Energy, the National Aeronautics and Space
Administration, the Japanese Monbukagakusho, the Max Planck Society, and the
Higher Education Funding Council for England. The SDSS Web Site is
http://www.sdss.org/.

The SDSS is managed by the Astrophysical Research Consortium for the
Participating Institutions. The Participating Institutions are the American
Museum of Natural History, Astrophysical Institute Potsdam, University of
Basel, University of Cambridge, Case Western Reserve University, University of
Chicago, Drexel University, Fermilab, the Institute for Advanced Study, the
Japan Participation Group, Johns Hopkins University, the Joint Institute for
Nuclear Astrophysics, the Kavli Institute for Particle Astrophysics and
Cosmology, the Korean Scientist Group, the Chinese Academy of Sciences
(LAMOST), Los Alamos National Laboratory, the Max-Planck-Institute for
Astronomy (MPIA), the Max-Planck-Institute for Astrophysics (MPA), New Mexico
State University, Ohio State University, University of Pittsburgh, University
of Portsmouth, Princeton University, the United States Naval Observatory, and
the University of Washington.

NM wishes to acknowledge Daria Dubinovska, Doroth\'ee Brauer,
Laura Hoppmann and the ESO service observers for assisting with the
observations, and helpful discussions with Professor Stephen Warren. The South
African SKA Project is acknowledged for 
funding the postdoctoral fellowship position at the University of Cape Town.
We thank the anonymous referee for helpful comments which improved
this paper.


\clearpage

\appendix

\section{Input Catalogue and Candidate List Completeness}\label{app:catcompl}

The appendix reviews the factors that reduce the effective sky-area of
the survey from the loss of objects due to factors mostly unrelated to the
intrinsic properties of the sources. 

\subsection{UKIDSS Catalogue Completeness}

The completeness of the UKIDSS-derived flux-limited stellar catalogue
was assessed by taking the SDSS DR7 quasar catalogue and
cross-matching to the NIR object sample. Less than 2~per~cent of the
quasars do not have counterparts with $K\le17.0$. Visual inspection of
the UKIDSS images demonstrates that the vast majority of the missing
sources are due to one of two reasons. The objects either fall in
small areas where the UKIDSS tiling is not contiguous (i.e. in small
areas of sky not covered by UKIDSS, the existence of which is already
incorporated in the UKIDSS survey area), or, possess $K$-band images
below the $K=17.0$ flux limit.

Just a handful of SDSS DR7 quasars which are clearly visible in the 
UKIDSS images remain but there is no photometry entry in the source 
catalogue. The objects represent a limitation to the completeness of
the UKIDSS photometric catalogue but the fraction of such images is
less than 0.2~per~cent and the effect on the survey area is negligibly
small. Thus the UKIDSS LAS catalogue is effectively 100~per~cent
complete with respect to quasars that appear in the SDSS. It is
assumed that the UKIDSS recovery rate of SDSS quasars is applicable to
all stellar sources.

\subsection{Catalogue Morphological Determination}\label{subsec:morphcompl}

The KX-selected quasar sample to be investigated is deliberately
restricted to include only objects with redshifts $z\ge1.0$. To
determine quantitatively whether such quasars may still be excluded
from the sample due to the presence of a host galaxy making them
appear non-stellar we investigated the fraction of SDSS quasars
classified as non-stellar compared to a control sample of
spectroscopically confirmed stars. 

Five-thousand randomly selected spectroscopically confirmed stars were
extracted from the SDSS database. Within the sample, 331
($\simeq$\,7~per~cent) are classified as morphologically extended by
the SDSS 
pipeline. Visual inspection of these objects reveal that all such
objects fall into one of three categories: i) close pairs of stellar
objects that have not been separated, ii) a superposition of a
foreground star and a background galaxy, or iii) the object is located
near a saturated star, which has compromised the photometric
parameters and hence the morphological parameters. All candidate
KX-selected quasars with such properties are removed in the final visual
inspection phase of the candidate list (Section~\ref{subsec:candsel}).
The experiment thus demonstrates that 7$\pm$1~per~cent of stellar
objects are `lost' due to morphological-classification issues in the
optical data. 

The same 5000 stars were also cross-matched to the UKIDSS database to
retrieve their NIR morphological classifications. Of the 1929 objects
that possess an entry in the database and also have $K\le16.6$
(i.e. the faint $K$-band limit of the current survey), 174
are classified as morphologically extended. These 174 objects would be
removed at the first stage of the candidate list construction. It is
therefore important to understand whether any of these objects were
erroneously removed, as it will affect the completeness of the final
`stellar'-catalogue. Of these 174 objects, 113 are also classed as 
extended by SDSS (see above). However, 61 ($\simeq$\,3\,per cent)
objects are classed as stellar by the SDSS pipeline. The majority are 
close pairs of objects where the SDSS image reduction has separated
the components but a single object is catalogued in the UKIDSS
catalogue\footnote{The reason is usually due to the presence of a
  relatively red component causing the two images to blend together in
  the UKIDSS images.}. Just five objects appear visually to not be
obviously extended, leading to an estimate of the fraction of stellar
images in the $K$-band sample lost due to erroneous morphological
classification by the UKIDSS piple line of just 5/1929, or 0.26\,per
cent. We assume that the same negligible correction factor applies to
the sample of morphologically stellar quasars.

To investigate the potential loss of quasars due to the host galaxy
causing the object to appear extended, the enhanced SDSS DR7 quasar catalogue was
searched for quasars with $z\ge1.0$ and extended morphological
classification. Just 268 of 78125 objects were found (0.34~per~cent)
and visual inspection of the SDSS images reveals they are virtually
all either close pairs of two point sources, a quasar with a low
redshift galaxy along the line of sight, or a quasar near a very
bright star which has affected the photometry. Thus, an insignificant
fraction of the quasar population at $z\ge 1.0$ is lost by 
requiring stellar morphologies at optical wavelengths.

However, the quasar host galaxy contributes more flux at longer
wavelengths, so the morphology as measured by UKIDSS is also
checked. Cross-matching the enhanced SDSS DR7 quasar catalogue to UKIDSS and
restricting to $z\ge1.0$ and $K\le16.6$, of 4324 matched objects that
are unresolved by the SDSS, 268 are classed as morphologically
extended in the NIR. Approximately half of these objects have
$z\le1.25$, indicating resolved host galaxy flux causing the NIR
images to be resolved. Of the 268, 30 are close pairs of objects and would
be removed by visual inspection, leaving 238/4324 = 5.5~per~cent of the known SDSS quasar
population at $z\ge1.0$ would be excluded at the first stage of candidate
selection.

\subsection{Photometry Defects}\label{subsec:photcompl}

A number of objects in the SDSS and UKIDSS catalogues are located
close to bright stars, or have another object located very close to the
line of sight. These objects have poor photometry and are removed from
consideration by visually inspecting the SDSS images of the candidates.

Within the 286.6 square degree CA region, 4.5~per~cent of the total
candidates suffer from such poor photometry. The percentages for the
NTT/VLT and NTT/VLT Deep regions are 6.2 and 4.9~per~cent, respectively.

\begin{table*}
\centering
  \caption{\label{tab:effarea}Reduction of effective area due to
    various effects.}
\begin{tabular}{cccc} \hline
 & CA & NTT/VLT & NTT/VLT Deep \\ 
 & (deg$^2$) & (deg$^2$) & (deg$^2$) \\ \hline \hline
Initial Area & 286.6 & 150.1 & 196.4 \\ \hline
Lost due to Stellar Requirement & -16.3 (5.7\%) & -8.6 (5.7\%) & -11.2 (5.7\%) \\
Remaining & 270.3 & 141.6 & 185.2 \\ \hline
Lost due to Bad Photometry & -12.2 (4.5\%) & -8.8 (6.2\%) & -9.1 (4.9\%)\\
Remaining & 258.1 & 132.8 & 176.1 \\ \hline \hline
Total & 567.0 & & \\ \hline
\end{tabular}
\end{table*}

\subsection{Effective Area}\label{subsec:eff_area}

The initial area of sky covered by the selection algorithm is 633.2
square degrees, but this is reduced by the loss of surveyed area due
to requiring the initial candidates to be unresolved and regions of sky
near bright stars that affect the photometry.
Table~\ref{tab:effarea} computes the loss of area due to
each of these factors, and the resulting effective area of the survey
of 567.0 square degrees.


\end{document}